\documentclass[journal,12pt,onecolumn,draftclsnofoot,romanappendices]{IEEEtran}

\IEEEoverridecommandlockouts

\usepackage{cite}
\usepackage{url}
\usepackage{enumerate}
\usepackage[dvipsnames]{xcolor}

\usepackage[T1]{fontenc}

\usepackage[cmex10]{amsmath}
\usepackage{amssymb}
\usepackage{amsthm}
\usepackage[makeroom]{cancel}
\usepackage{booktabs}
\usepackage[ruled,linesnumbered,vlined]{algorithm2e}

\makeatletter
\newcommand{\nosemic}{\renewcommand{\@endalgocfline}{\relax}}
\makeatother
\usepackage{cryptocode}[=2018-11-11]

\usepackage{multirow}
\usepackage{diagbox}

\usepackage{balance}

\usepackage{todonotes}

\usepackage{tikz}
\usetikzlibrary{matrix,arrows,shapes,positioning,chains,scopes,decorations.markings,calc,patterns}
\tikzset{->-/.style={decoration={markings,mark=at position .5 with {\arrow{>}}},postaction={decorate}}}
\usepackage{pgfplots}
\usepgfplotslibrary{groupplots}
\usetikzlibrary{pgfplots.groupplots}

\theoremstyle{definition}

\DeclareMathOperator{\slip}{slip}
\DeclareMathOperator{\eff}{Eff}
\DeclareMathOperator{\pe}{PE}

\newcommand{\effpe}{\eff^\text{AllPairs}}
\newcommand{\effu}{\eff^\text{Neighb}}
\newcommand{\evar}{\times 10^{-5}}

\makeatletter
\newcommand{\removelatexerror}{\let\@latex@error\@gobble}
\def\pgfplots@getautoplotspec into#1{%
\begingroup
\let#1=\pgfutil@empty
\pgfkeysgetvalue{/pgfplots/cycle multi list/@dim}\pgfplots@cycle@dim
\let\pgfplots@listindex=\pgfplots@numplots
\pgfkeysgetvalue{/pgfplots/cycle list set}\pgfplots@listindex@set
\ifx\pgfplots@listindex@set\pgfutil@empty
\else 
\c@pgf@counta=\pgfplots@listindex
\c@pgf@countb=\pgfplots@listindex@set
\advance\c@pgf@countb by -\c@pgf@counta
\globaldefs=1\relax
\edef\setshift{%
\noexpand\pgfkeys{
/pgfplots/cycle list shift=\the\c@pgf@countb,
/pgfplots/cycle list set=
}
}%
\setshift%
\fi
\pgfkeysgetvalue{/pgfplots/cycle list shift}\pgfplots@listindex@shift
\ifx\pgfplots@listindex@shift\pgfutil@empty
\else
\c@pgf@counta=\pgfplots@listindex\relax
\advance\c@pgf@counta by\pgfplots@listindex@shift\relax
\ifnum\c@pgf@counta<0
\c@pgf@counta=-\c@pgf@counta
\fi
\edef\pgfplots@listindex{\the\c@pgf@counta}%
\fi
\ifnum\pgfplots@cycle@dim>0
\c@pgf@counta=\pgfplots@cycle@dim\relax
\c@pgf@countb=\pgfplots@listindex\relax
\advance\c@pgf@counta by-1
\pgfplotsloop{%
\ifnum\c@pgf@counta<0
\pgfplotsloopcontinuefalse
\else
\pgfplotsloopcontinuetrue
\fi
}{%
\pgfkeysgetvalue{/pgfplots/cycle multi list/@N\the\c@pgf@counta}\pgfplots@cycle@N
\pgfplotsmathmodint{\c@pgf@countb}{\pgfplots@cycle@N}%
\divide\c@pgf@countb by \pgfplots@cycle@N\relax
\expandafter\pgfplots@getautoplotspec@
\csname pgfp@cyclist@/pgfplots/cycle multi list/@list\the\c@pgf@counta @\endcsname
{\pgfplots@cycle@N}%
{\pgfmathresult}%
\t@pgfplots@toka=\expandafter{#1,}%
\t@pgfplots@tokb=\expandafter{\pgfplotsretval}%
\edef#1{\the\t@pgfplots@toka\the\t@pgfplots@tokb}%
\advance\c@pgf@counta by-1
}%
\else
\pgfplotslistsize\autoplotspeclist\to\c@pgf@countd
\pgfplots@getautoplotspec@{\autoplotspeclist}{\c@pgf@countd}{\pgfplots@listindex}%
\let#1=\pgfplotsretval
\fi
\pgfmath@smuggleone#1%
\endgroup
}

\pgfplotsset{
cycle list set/.initial=
}
\makeatother

\begin{document}

\title{Intrablock Interleaving for Batched Network Coding with Blockwise Adaptive Recoding}
\author{\IEEEauthorblockN{Hoover~H.~F.~Yin, Ka~Hei~Ng, Allen~Z.~Zhong, Raymond~W.~Yeung, Shenghao~Yang, and Ian~Y.~Y.~Chan}
	\thanks{This paper was presented in part at 2021 IEEE International Symposium on Information Theory \cite{intrablock}.}
	\thanks{H.~Yin is with the n-hop technologies Limited, Hong Kong, China and the Institute of Network Coding, The Chinese University of Hong Kong, Hong Kong, China.
	R.~Yeung is with the same institute and also with the Department of Information Engineering, The Chinese University of Hong Kong, Hong Kong, China.
	He is also a Principal Investigator of the Centre for Perceptual and Interactive Intelligence (CPII) Limited.
	K.~Ng is with the Department of Physics, The Chinese University of Hong Kong, Hong Kong, China.
	A.~Zhong is with the Department of Computer Science and Engineering, The Chinese University of Hong Kong, Hong Kong, China.
	S.~Yang is with the School of Science and Engineering, The Chinese University of Hong Kong, Shenzhen, Shenzhen, China.
	He is also with Shenzhen Key Laboratory of IoT Intelligent Systems and Wireless Network Technology and Shenzhen Research Institute of Big Data, Shenzhen, China.
	I.~Chan is with the Department of Economics, The Chinese University of Hong Kong, Hong Kong, China.
	Emails: \mbox{hfyin@inc.cuhk.edu.hk}, \mbox{whyeung@ie.cuhk.edu.hk}, \mbox{kaheicanaan@link.cuhk.edu.hk}, \mbox{zwzhong@cse.cuhk.edu.hk}, \mbox{shyang@cuhk.edu.cn}, \mbox{chanyy@link.cuhk.edu.hk}
	}
	\thanks{This work was funded in part by the Shenzhen Science and Technology Innovation Committee (Grant JCYJ20180508162604311, ZDSYS20170725140921348).}
}

\maketitle

\begin{abstract}
Batched network coding (BNC) is a low-complexity solution to network transmission in multi-hop packet networks with packet loss.
BNC encodes the source data into batches of packets.
As a network coding scheme, the intermediate nodes perform recoding on the received packets belonging to the same batch instead of just forwarding them.
A recoding scheme that may generate more recoded packets for batches of a higher rank is also called adaptive recoding. Meanwhile, in order to combat burst packet loss, the transmission of a block of batches can be interleaved.  Stream interleaving studied in literature achieves the maximum separation among any two consecutive packets of a batch, but permutes packets across blocks and hence cannot bound the buffer size and the latency. 
To resolve the issue of stream interleaver, we design an intrablock interleaver for adaptive recoding that can preserve the advantages of using a block interleaver when the number of recoded packets is the same for all batches.
We use potential energy in classical mechanics to measure the performance of an interleaver, and propose an algorithm to optimize the interleaver with this performance measure.
Our problem formulation and algorithm for intrablock interleaving are also of independent interest.
\end{abstract}

\section{Introduction}

In the era of Internet-of-Things (IoT), multi-hop wireless networks become popular in smart cities applications.
Unlike wired links, wireless links are not reliable as they are easily interfered by other wireless signals and environment factors.
In particular, packet loss, especially burst loss, is a common phenomenon at each wireless link.
Traditional networking approaches based on forwarding and end-to-end retransmission do not perform well in wireless multi-hop networks because a packet can reach its destination only if it is transmitted successfully at all the links, whose probability diminishes exponentially fast with the number of hops.

Linear network coding \cite{flow,alg,linear} is one of the solutions for reliable communication in wireless multi-hop networks.
\emph{Random linear network coding (RLNC)} \cite{random,random2,jaggi03,Sanders03} is a realization of network coding which can be applied without depending on feedback or knowledge of the network topology.
Instead of forwarding, the intermediate nodes transmit packets generated by random linear combinations of the received packets, also called \emph{recoding}.
In generation-based RLNC~\cite{chou03}, however, the encoding/decoding computational cost and the coefficient vector overhead may prevent practical implementation if the number of packets in a generation, i.e., the \emph{generation size}, is large.

\emph{Batched network coding (BNC)} \cite{Silva2009,Heidarzadeh2010,Mahdaviani12,yang14bats} is a variation of RLNC which can resolve the above issues by using an inner-code-outer-code structure.
For a relatively large generation of packets for transmission, the outer code encodes the packets of the generation into relatively small subsets of packets called \emph{batches} (also known as chunks and classes).
The inner code is formed by recoding in a batch-by-batch manner at all the network nodes.
The number of packets in each batch generated by the outer code is called the \emph{batch size}.
The coefficient vector overhead and the recoding computational and storage costs all depend on the batch size, but not the generation size.\footnote{As an intermediate node does not need to perform coding crossing batches, if all the recoded packets of one batch have been transmitted, the batch can be discarded.} 
When there is only a single batch, BNC becomes the generation-based RLNC, or its variations~\cite{lucani18fulcrum,nguyen20dsep} which use binary field in the inner code to reduce the computational cost.

In this paper, we focus on BNC with multiple batches, that can be generated for example by overlapping subsets of packets \cite{Silva2009,Heidarzadeh2010,yaoli11,bin_expander15}, or by extending fountain codes and LDPC codes~\cite{yang14bats,bin18ldpc}.
For such batched network codes, the decoding of batches can help each other so that is it not necessary that each batch can be decoded by itself.
The achievable rate of BNC is upper bounded by the average rank of the end-to-end batch transfer matrices, which can be achieved by random linear outer codes~\cite{yang11x2}.
There exists batched network code which has close-to-optimal achievable rate, say, BATS codes \cite{yang14bats,bats_book}, where the outer code is a matrix extension of Raptor codes~\cite{shokRaptor,Raptormono}.
For BATS codes, the outer code encoding and decoding complexities per packet depend on the batch size, but not the generation size. In contrast, for generation-based RLNC, even encoding and recoding can be sparse and have low computational cost (such as DSEP Fulcrum~\cite{nguyen20dsep}), the decoding computation cost still increases fast with the generation size.

When the batch size tends to infinity, the achievable rate of BNC tends to the min-cut of the network in a general setting~\cite{Dana2006,Lun2008}.
In practice, a small batch size, e.g., $8$ or $16$, not only achieves a good rate when the packet losses are independent, but also incurs a small coefficient vector overhead, a small computational cost and a small buffer requirement~\cite{bats_book}.
Moreover, a small batch size makes latency of recoding manageable even for dense recoding. In contrast, for generation-based RLNC, dense recoding may incur significant delay as a number of packets close to the generation size must be collected before recoding, and the delay accumulates hop by hop.

There is another line of works about RLNC that employ feedback of packet reception to adjust the coding behaviors~\cite{sundararajan2008arq,sundararajan2011network,cohen2020adaptive}. In this paper, we do not explicitly use feedback of packet reception, but recoding at a network node may need the packet loss statistics of its adjacent links, which may need feedback to acquire. If the channel statistics is stable, we may not need to have frequently feedback to update the packet loss statistics. Therefore, the BNC schemes we discuss apply to the communication scenarios where feedback has long delay and is not reliable.

\subsection{Paper Motivation}

Various studies have revealed that burst packet loss, which is common in wireless communication, degrades
the throughput of BNC with a relatively small batch size~\cite{protocol,yang14a,ge_adaptive}.
Although we can improve the throughput by considering burst loss models when optimizing the number of recoded packets per batch \cite{ge_adaptive,yin21impact}, the throughput can be significantly enhanced by interleaving \cite{protocol,recoding}.
In a traditional forwarding scenario, the source node interleaves the packets and the destination node deinterleaves the packets, while the packets are kept in an interleaved order at the intermediate nodes.
For BNC with multiple batches, recoding is performed on the received batches at each node.
Technically, the batches are deinterleaved and then reinterleaved.
Therefore, the interleaving techniques for BNC should be designed together with recoding.

Two recoding-interleaving approaches for BNC have been proposed %
in the literature.
The first approach is called \emph{Baseline Recoding and Block Interleaving (BR-BI)}.
A number of consecutively generated batches are grouped together as a \emph{batch block}.
The main difference between a batch block and a generation is that we do not need to decode the batch blocks one by one.
That is, the BNC jointly decodes multiple batch blocks so that even when many packets in a batch block are lost, the remaining packets can still contribute the decoding of BNC.
The \emph{baseline recoding scheme} generates the same number of recoded packets for every batch regardless of the number of received packets of the batches, which has been used widely in the analysis of BNC~\cite{dong20,fun,zhou17b,bats_schedule,delay,buffer}.
With baseline recoding, a \emph{block interleaver} \cite{block_interleaver,interleaver} can be applied to transmitted the packets of the batches in a block in a round-robin manner \cite{pro2}.
For baseline recoding, the block interleaver achieves \emph{perfect interleaving} in the sense that every pair of consecutive packets of each batch is separated by exactly the number of packets of other batches in the block.
For example, if we group the first $6$ batches, each of $4$ packets, into $2$ batch blocks, each of $3$ batches,
we obtain the following sequence of packet transmission
\begin{equation*} \label{eq:seq_block}
	\mathtt{\textcolor{red}{123123123123}\textcolor{blue}{456456456456}}\ldots
\end{equation*}
where the $i$-th number $s$ in the sequence means that the $i$-th transmitted packet belongs to the $s$-th batch.
In the above example, the first $12$ transmissions belong to the first batch block.
That is, we have a clear block boundary to distinguish different batch blocks.
In other words, a batch block occupies certain number of consecutive transmissions.

The second approach is called \emph{Adaptive Recoding and Stream Interleaving (AR-SI)}.
Unlike baseline recoding, \emph{adaptive recoding} allows different number of recoded packets to be transmitted for different batches.\footnote{The adaptive recoding here does not need the feedback about the reception of the previously transmitted packet of the batch. Existing adaptive recoding algorithms use only the ranks of batches, which can be calculated from the coefficient vectors, and the link packet loss statistics.}
In general, a higher number of recoded packets tends to be transmitted for a batch of a higher rank.
Adaptive recoding can achieve a higher expected rank of batch transfer matrices than baseline recoding under the same network link resource~\cite{ge_adaptive,rf,scheduling,adaptive}.
However, with adaptive recoding, interleaving within a batch block (which occupies consecutive transmissions) cannot achieve perfect interleaving in general. %
A \emph{stream interleaver} was proposed to achieve perfect interleaving for adaptive recoding \cite{protocol}, where the packets of the batches %
are organized into multiple interleaving streams in a way that the packets of the same batch are assigned to the same stream.
The packets from these streams are transmitted in a round-robin manner.
  The following example illustrates a stream interleaver of $3$ streams: 
  \begin{equation*} \label{eq:seq_perfect}
	\begin{matrix}
		\mathtt{\textcolor{red}{11111}\textcolor{blue}{6}}\ldots\\
		\mathtt{\textcolor{red}{22}\textcolor{blue}{4444}}\ldots\\
		\mathtt{\textcolor{red}{333}\textcolor{blue}{555}}\ldots
	\end{matrix} \longrightarrow	\mathtt{\textcolor{red}{1231231}\textcolor{blue}{4}\textcolor{red}{31}\textcolor{blue}{45}\textcolor{red}{1}\textcolor{blue}{45645}}\ldots
\end{equation*}

Though achieving the same interleaving effect but better recoding efficiency than BR-BI, AR-SI introduces an implementation issue. %
Due to causality or the effectiveness of recoded packets, we may need to introduce some idle timeslots in order to maintain the distance of separation and the structure of the interleaver.
As we do not have a natural boundary of the transmission of a batch, i.e., the time interval from the first packet to the last packet of a batch is not bounded, maintaining the structure of the interleaver tends to increase the latency.
Moreover, the buffer size is not bounded due to this reason.

In contrast, we have deterministic borders of the batch blocks so that the latency and buffer size are under control.
For example, a node only has to store a batch block for the packets which are transmitting, and also a batch block for the incoming packets.
This is because after recoding and transmitting the packets of a batch, the batch can be discarded as it does not contribute to the inner code of other batches.
Also, as we perform recoding in a block-by-block manner (blockwisely), the latency is bounded at most for one batch block.
Generally speaking, as long as the average number of incoming batches and outgoing batches are conserve, the latency and buffer size are bounded even when the incoming and outgoing batch blocks have different number of batches in it \cite{delay,buffer}.

\subsection{Paper Contributions and Organization}

On one hand, recoding blockwisely does not incur the latency and buffer size issues as in AR-SI.
On the other hand, adaptive recoding can outperform baseline recoding but unlike BR-BI, a standard block interleaver does not support adaptive recoding as the number of recoded packets for each batch may not be the same.
To combine the advantages of adaptive recoding and interleaving, we are interested in an Adaptive Recoding and IntraBlock Interleaving (AR-IBI) approach which 
groups certain batches into a batch block, applies adaptive recoding to decide the number of recoded packets for these batches, and tries to separate the packets belonging to the same batch as far as we can inside the block.
We call the interleaving problem of a block with batches of different numbers of packets the  \emph{intrablock interleaving} problem.

We propose a joint intrablock interleaving and adaptive recoding optimization problem, together with a heuristic decomposition, where the intrablock interleaving optimization problem and the adaptive recoding optimization are solved iteratively.
As we cannot guarantee perfect interleaving, a general performance measure of intrablock interleaving is desired.
Motivated by some physics phenomenons, we borrow a concept from classical mechanics called \emph{potential energy} to measure the performance of intrablock interleaving, which also has an interpretation in economics as utility.
We propose a general approximation scheme to efficiently produce an intrablock interleaver which acts as a good starting point for other methods to further improve the interleaver.
We also propose a tuning scheme for this purpose.

The remainder of the this paper is organized as follows. In Section~\ref{sec:sys}, we introduce the network model and the batched network coding scheme. In Section~\ref{sec:sribi}, we introduce adaptive recoding and intrablock interleaving, and then discuss the joint optimization problem of the both, i.e., AR-IBI.
Next, we introduce our performance metric of intrablock interleaving in Section~\ref{sec:model} and discuss an approximation scheme to optimize the intrablock interleaving problem in Section~\ref{sec:solver}.
Then, we evaluate the performance of AR-IBI numerically in Section~\ref{sec:sim}.
Lastly, we conclude the paper in Section~\ref{sec:conclude}.

\section{Network Model and Batched Network Coding}
\label{sec:sys}

In this section, we first introduce how to transmit a file in a line topology network where  packet loss can occur at the links in the network. The discussion of this paper can be extended into some more general network topologies as in~\cite{bats_book}.

\subsection{Network Model}

\begin{figure}
	\centering
	\begin{tikzpicture}[font=\footnotesize,dot/.style={circle,draw,thick,inner
		sep=1pt,minimum size=7pt}]
		\node[dot] (s) at(-2,0) {};
		\node[dot] (a1) at(0,0) {} edge[<-] (s);
		\node[dot] (a2) at(2,0) {} edge[<-] (a1);
		\node[dot] (t) at(4,0) {} edge[<-] (a2);
	\end{tikzpicture}
	\caption{A three-hop line network where network links only exist between two neighboring nodes.
	}
	\label{fig:line}
\end{figure}
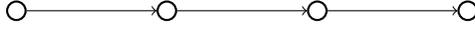

We consider a multi-hop line network formed by a finite sequence of network nodes, where the first node is the source node and the last node is the destination node. All the other nodes are called the intermediate nodes. Network links exist only between two consecutive network nodes. 
Fig.~\ref{fig:line} illustrates an example of a three-hop line network.

We assume each network link is a packet erasure channel, where a packet transmitted through the link can be either correctly received or completely erased.
Assume time is real. A packet can be transmitted at any time, but there must be at least a unit time interval between two consecutively transmitted packets. This assumption says that the bandwidth of each link is $1$ packet per unit time. For a number of packets transmitted through a network link, the packet loss pattern of these packets is assumed to be \emph{stationary}. Specifically, suppose totally $k$ packets are transmitted, where the $i$-th packet is transmitted at time $\tau_i$. Let $Z_{\tau_i}$ be the indicator random variable specifying whether the $i$-th packet is correctly received. Then, for any real number $\tau$, $(Z_{\tau_i})_{i=1}^k$ and $(Z_{\tau_i+\tau})_{i=1}^k$ have the same distribution.  

Except for the stationary condition, our design to be discussed does not restrict to a specific packet loss model. We will discuss some specific packet loss models in Section~\ref{sec:lossmodel} for the purpose of numerical evaluation. In particular, we will use a continuous time version of the Gilbert-Elliott (GE) model \cite{GilbertBurst,ElliottBurst} in our simulation.

\subsection{Batched Network Coding}

Fix a finite field $\mathbb{F}$ and a positive integer $M$.
The file to be transmitted is divided into multiple \emph{input packets} of equal length.
Each input packet is regarded as a column vector over $\mathbb{F}$.
We abuse the notation to denote a set of packets as a matrix formed by juxtaposing the packets in the set. A BNC scheme includes an outer code and an inner code. 

The source node applies an outer code encoder of a BNC scheme to generate a sequence of \emph{batches}, each of which consists of $M$ coded packets generated from the input packets.
There are different methods in the literature to generate batches. Our interleaving problem does not dependent on a specific outer code. 
Here, $M$ is also known as the \emph{batch size}.
A coefficient vector of $M$ symbols from $\mathbb{F}$ is attached to each coded packet in a batch, and the coefficient vectors of all the $M$ packets of a batch form an identity matrix.

The inner code is formed by \emph{recoding} at the network nodes. %
A network node starts recoding a batch after all the packets of the batch which are not lost at the previous link are received. Recoded packets of a batch are the random linear combinations of the received packets of the batch, and the coefficient vectors of the recoded packets are the same linear combinations of the corresponding coefficient vectors.
The network node should decide the number of recoded packets to be generated for a batch and the sequence of recoded packets for transmitting to the next node, which will be further discussed in the following parts of this section. 

For a batch received at a network node, two packets of the batch are called \emph{linearly independent} if and only if their coefficient vectors are linearly independent.
The rank of the matrix formed by the coefficient vectors of all the packets of the batch is also called the \emph{rank} of the batch, which measures the information carried by the batch.
At the destination node, all the batches are decoded jointly, thus it is not necessary that the rank of a batch is sufficiently large so that the batch can be solved. 
A necessary condition such that all the input packets can be decoded is that the total rank of all the received batches is at least the number of input packets.
For the outer codes introduced in \cite{yang14bats,bin_expander15,bin18ldpc}, a belief propagation (BP) algorithm can be applied to decode the batches efficiently and achieve a rate very close to the average rank of all the batches used for decoding.

\section{Adaptive Recoding and Intrablock Interleaving}
\label{sec:sribi}

In this section, we formulate \emph{adaptive recoding} and \emph{intrablock interleaving}, and discuss the joint optimization problem of them.

\subsection{Recoding and Interleaving}

Let $L$ be a positive integer called \emph{block size}. Every $L$ batches generated at the source node are grouped together, called a \emph{batch block}.
For simplicity, we also call it a block.
We adopt the adaptive recoding scheme in \cite{adaptive,yin21impact}, which perform recoding in a block-by-block manner.
Consider a block of $L$ batches where the $k$-th batch has $t_k$ recoded packets to transmit. The total number of packets in the block, $T := \sum_{i=1}^Lt_k$, is a fixed integer that can be determined by resource allocating~\cite{dong20}. In this paper, we assume that $T$ is given, and $t_k$ can be optimized using the ranks of the batches in the block together with the packet loss model.
we will discuss how to optimize $t_k$ together with interleaving in the next subsection.
As this adaptive recoding approach considers a block of batches, it is also called \emph{blockwise adaptive recoding}. %

Assume one packet is transmitted per unit time.
Let $\mathcal{M} = \{1, 2, \ldots, T\}$ be the set of time for sending the packets in a block. 
Define a mapping $s \colon \mathcal{M} \to \{1, 2, \ldots, L\}$, where %
$s(i) = k$ means that a packet belonging to the $k$-th batch is sent at time $i$. %
We call such a mapping $s$ a \emph{transmission sequence}, and write $s$ as the sequence $s(1) s(2) \ldots s(T)$.
See an example of transmission sequence for intrablock interleaving with a block of $3$ batches:
\begin{equation} \label{eq:seq_intra}
  1312312131
\end{equation}

The preimage $s^{-1}(k) \subseteq \mathcal{M}$ is the set of time at which the packets of the $k$-th batch are transmitted.
Given $\{t_k\}_{k = 1}^L$, the set of all valid $s$ is denoted by $\mathcal{F}(\{t_k\}_{k = 1}^L)$.
In the following of this section, we discuss the optimization of $\{t_k\}_{k = 1}^L$ and $s$ for this scheme.

\subsection{Joint Optimization}

At the source node, a newly generated batch has rank $M$.
Due to packet loss, the rank of a batch can only decrease when it passes through the network.

Let $r_k$ be the rank of the $k$-th batch in the block at a network node, which can be calculated from the coefficient vectors of the batch. %
Note that we can generate more than $r_k$ recoded packets for this batch but it is not possible to increase the rank of this batch.
Recall that $Z_\tau$ is the indicator random variable specifying whether the packet sent at time $\tau$ is correctly received.
The expected rank of the $k$-th batch at the next node when we transmit its packets following a transmission sequence $s$ is  %
\begin{equation*} %
  E(r_k, s^{-1}(k)) := \sum_{i = 0}^{t_k} \Pr\left( \sum_{\tau\in s^{-1}(k)}  Z_{\tau} = i\right) R(i,r_k),
\end{equation*}
where %
$R(i,r_k)$ is the expected dimension of the vector space spanned by $i$ random vectors sampled from an $r_k$-dimensional vector space over $\mathbb{F}$.
The function $E(\cdot,\cdot)$ is called the \emph{expected rank function}.

Note that $\Pr(\sum_{\tau\in s^{-1}(k)}  Z_{\tau} = i)$ depends specifically on $s^{-1}(k)$.
We leave the evaluation of $\Pr(\sum_{\tau\in s^{-1}(k)}  Z_{\tau} = i)$ for GE models to Appendix~\ref{sec:ge:exp} as an example.
Though we have an exact formula of $R(i,r)$ in \cite{yang14bats}, we can apply the approximation $R(i,r) \approx \min\{i,r\}$ to reduce the computation cost when the field size is sufficiently large, say, $|\mathbb{F}| = 2^8$.
The effectiveness of the approximation has been verified in  \cite{adaptive,ge_adaptive,coupon,field_size}.

The objective of blockwise adaptive recoding optimization is to maximize the average expected rank of the batches at the next node~\cite{adaptive,yin21impact}.
Here, we extend the optimization to include the transmission sequence as a variable:
\begin{equation}
  \tag{BAR} \label{eq:BAR}
  \max_{t_1,\ldots,t_L, s} \frac{1}{L} \sum_{k = 1}^L E(r_k, s^{-1}(k))\quad \text{s.t.} \quad \sum_{k = 1}^L t_k = T \text{ and } s\in \mathcal{F}(\{t_k\}_{k=1}^L).
\end{equation}
Here, BAR stands for Blockwise Adaptive Recoding.
Note that \eqref{eq:BAR} is a combinatorial optimization problem where the size of the search space $\sum_{t_1,\ldots,t_L:\sum_{k = 1}^L t_k = T} \binom{T}{t_1, t_2, \ldots, t_L}$ is exponentially large.
To tackle this problem, we propose a heuristic approach to decompose the problem.
First, given a transmission sequence $s$, we approximate the effect of interleaving and update $\{t_k\}_{k = 1}^L$.
Next, given $\{t_k\}_{k = 1}^L$, we optimize the intrablock interleaver to update the transmission sequence $s$.
We iterate the above two steps for a few rounds and select the transmission sequence which can achieve the highest expected rank at the next node.
We observe in our numerical evaluation that we only need one or two iterations of the above two steps. %
We present the details of these two steps in the remaining text of this section.

\subsection{First Step in the Decomposition}

In BR-BI, the interleaver depth of the block interleaver is the packet separation, i.e., the number of packets separated for two consecutive packets of the same batch.
Although we want to separate the packets of the same batch uniformly to prevent bias, we may not be able to do so in an intrablock interleaver.
To measure the ideal separation of a batch in a transmission sequence, we define the \emph{pseudo interleaver depth} as the average packet separation.
We can see that the pseudo interleaver depth of a block interleaver is equivalent to the interleaver depth.

As a formal definition, write
\begin{equation} \label{eq:s^-1}
	s^{-1}(k) = \{x_{k,1}, x_{k,2}, \ldots, x_{k,t_k}\}
\end{equation}
where $x_{k,1} < x_{k,2} < \ldots < x_{k,t_k}$.
The \emph{pseudo interleaver depth} of the $k$-th batch, denoted by $L_k$, is defined as
\begin{equation*}
	L_k := \begin{cases}
		\frac{1}{t_k-1} \sum_{j=1}^{t_k-1}(x_{k,j+1}-x_{k,j}) = \frac{x_{k,t_k}-x_{k,1}}{t_k-1} & \text{if } t_k > 1,\\
		1 & \text{otherwise}.
	\end{cases}
\end{equation*}
We evaluate the pseudo interleaver depths of GE models in Appendix~\ref{sec:ge:pid}.

We now describe the first step of the decomposition.
Fix a transmission sequence $s$.
We consider the pseudo interleaver depth of each batch as the packet separation to express the packet loss pattern for the batch induced by $s$.
At the beginning, we do not have a transmission sequence for us to calculate the pseudo interleaver depth.
We can set $L_k = 1$ for all $k$, which represent the case that the packets of the batches are sent without interleaving.

Without loss of generality, we can consider the first packet of a batch is sent at time $1$ due to stationarity.
According to the pseudo interleaver depth, the $t_k$ packets of the $k$-th batch are sent at time indicated in the set $\mathcal{S}_{L_k}(t_k)$ where
\begin{equation*}
	\mathcal{S}_\ell(t) := \begin{cases}
		\{1, \ell+1, 2\ell+1, \ldots, (t-1)\ell+1\} & \text{for positive integers $t$},\\
		\emptyset & \text{otherwise}.
	\end{cases}
\end{equation*}
As $L_k$ may not be an integer, $\mathcal{S}_{L_k}(t_k)$ may include non-integers.
Note that we allow time to be real, so $E(r_k, \mathcal{S}_{L_k}(t_k))$ is still well-defined. 

With the updated description of expected rank functions, we can formulate the following blockwise adaptive recoding problem which is similar to that modeled in \cite{adaptive,yin21impact}:
\begin{equation*}
	\max_{t_1,\ldots,t_L} \frac{1}{L} \sum_{k = 1}^L E(r_k, \mathcal{S}_{L_k}(t_k)) \quad \text{s.t.} \quad \sum_{k = 1}^L t_k = T.
\end{equation*}
As the packet loss pattern for each batch is stationary, the expected rank function $E(r,\mathcal{S}_{L_k}(t_k))$ is monotonically increasing and concave with respect to $t_k$ \cite{uni}.
This concavity allows us to solve the above problem efficiently by greedy algorithms similar to that proposed in \cite{adaptive,uni}.
After we obtain the set of $\{t_k\}_{k = 1}^L$ solving the above optimization problem, we pass it to the second step as described as follows.

\subsection{Second Step in the Decomposition}

In the second step, we fix the number of recoded packets $\{t_k\}_{k = 1}^L$.
We aim to find a proper $s\in \mathcal{F}(\{t_k\}_{k = 1}^L)$, a feasible solution of \eqref{eq:BAR}. The spirit of interleaving is to separate the packets belong to the same batch as far as possible. When $t_k$ are the same for all $k$, we know that  the optimal solution is to evenly separate two consecutive packets of a batch by $L$, which is already the largest possible separation according to our definition of transmission sequence.

For the general case that $t_k$ may be different, we can formulate an optimization problem 
\begin{equation} \tag{DE} \label{eq:DE}
  \max_{s \in \mathcal{F}(\{t_k\}_{k = 1}^L)} \eff(s)
\end{equation}
where $\eff(s)$ is a certain measure of $s$, called the \emph{dispersion efficiency}.
Motivated by \emph{potential energy} used in classical mechanics (to be further explained in Section~\ref{sec:model}), we 
define two types of dispersion efficiencies for some function $g \colon \mathbb{R}^+ \to \mathbb{R}$:
\begin{IEEEeqnarray*}{rCl}
	\effpe_g (s) & := & \sum_{k = 1}^L \sum_{i = 1}^{t_k-1} \sum_{j = i+1}^{t_k} g(x_{k,j}-x_{k,i}), \label{eq:ap}\\
	\effu_g (s) & := & \sum_{k = 1}^L \sum_{i = 1}^{t_k-1} g(x_{k,i+1}-x_{k,i}), \label{eq:nb}
\end{IEEEeqnarray*}
where the variables $x_{k,i}$ are elements in $s^{-1}(k)$ as denoted
in \eqref{eq:s^-1}. 

Let us give some further explanation about the two types of dispersion efficiencies. 
The dispersion efficiency $\effpe_g$ considers the distances between all pairs of packets in one batch, and dispersion efficiency $\effu_g$ considers only the distances between two consecutive packets in one batch.

Examples of $g(x)$ are $-\frac{1}{x}$, $-\frac{1}{x^2}$, $\ln(x)$ and $\tan^{-1}(x)$. These functions are all increasing and concave, and can be used as a \emph{utility function} in economics to measure the welfare of a consumer as a function of the consumption of resources. Moreover, $\ln(x)$ and $\tan^{-1}(x)$ are used as the utility functions to analyze the congestion control algorithms of TCP Vegas and TCP Reno, respectively \cite{low2017analytical}.

We will discuss the aforementioned measure in Section~\ref{sec:model} and the corresponding algorithms to optimize \eqref{eq:DE} in Section~\ref{sec:solver}.

\section{Intrablock Interleaving Performance Measures} %
\label{sec:model}

A pair of particles (or objects, charges, etc.) has a force exerting on each other, either attractive or repulsive.
In real world scenarios, the longer the distance between two particles, the weaker the interaction between these particles.
In other words, if the interaction between two particles is repulsive, then the repulsive force is stronger when the two particles are closer.
This characteristic suits our desired property of a good intrablock interleaver.
The idea is that we regard a packet as a particle.
The ``repulsion effect'' between two packets belonging to the same batch is stronger when they are being put too close in a transmission sequence.

In classical mechanics, the potential energy associated with an applied force on a particle is the work done against that force to move the particle from a reference point to the current position.
In other words, a particle has a potential energy due to the interaction with another particle.
The force, which is a vector, is the negative gradient of the potential energy.
Conversely speaking, we can derive the potential energy by integrating the force in the opposite direction on a trajectory.
An important nature, known as the \emph{minimum total potential energy principle} \cite{minPE}, is that each particle tries to attain the lowest potential energy.
It is the principle of least action \cite{feynman2} in an equilibrium state when the kinetic energy is zero.

We take electrostatic potential \cite{electrostatic} as an example.
Suppose we fix the locations of two identical charges. %
If we put another identical charge between the two fixed charges colinearly and assume no other external force, e.g., friction, acting on the charge, then each of the fixed charges repulses the new charge to the direction of another fixed charge.
At the end, the equilibrium is that the new charge is located at the mid-point between the two fixed charges and the potential energy of this charge attains its minimum.
The scenario becomes complicated when we introduce more particles because each particle interacts with all the other particles.
Due to the fact that the interaction of two particles is weaker when the distance is longer, the Ising model \cite{ising} in statistical mechanics suggests that we can approximate the total potential energy by considering the neighboring particles only.
This way, minimizing the total potential energy would separate the particles as far as possible without biasing towards some particles when all the particles have the same physical quantity (charge, mass, etc.).
We give a detailed discussion regarding this phenomenon in Appendix~\ref{sec:potential}.
In short, we can model a potential energy minimization problem to obtain an optimal transmission sequence.

\subsection{Potential Energy Models}
\label{sec:pe_model}

We briefly describe below some commonly used potential energy models for the interaction between particles. %

\subsubsection{Newtonian Gravitational Potential}

One of the most well-known forces between two particles is the gravitation force \cite{gravitation}.
In classical mechanics, the \emph{Newtonian gravitational potential energy} of a point mass $m_1$ in the present of another point mass $m_2$ is given by $-G \frac{m_1 m_2}{r}$, where $G$ is the gravitational constant and $r$ is the distance between the two point masses.
The negativity of the potential energy means that the gravitational force is an attractive force.
By applying a transformation via time reversal symmetry, 
the force becomes repulsive so that %
the potential energy becomes positive.

\subsubsection{Electrostatic Potential}

In electromagnetism, the \emph{electrostatic potential energy} \cite{electrostatic} of a point charge $q_1$ in the present of another point charge $q_2$ is $k_e \frac{q_1 q_2}{r}$ where $k_e$ is the Coulomb's constant and $r$ is the distance between these two point charges.
When the two charges have the same sign, the potential energy is positive, i.e., the force acting on them is repulsive.

\subsubsection{Lennard-Jones Potential}
In atomic physics, two neutral atoms repulse each other when the distance between them is too short, while they attract each other when the distance between them is long enough.
The \emph{Lennard-Jones potential (L-J potential)} \cite{LJ} is a common formulation of the intermolecular potential 
of two atoms which has the general form of $\frac{A}{r^n} - \frac{B}{r^m}$ for some $A, B \ge 0$ and $n, m > 0$, where $r$ is the distance between the two atoms.\footnote{%
When we describe the L-J potential between two neutral atoms, 
we have $A = 4\epsilon \sigma^{12}$, $B = 4\epsilon \sigma^6$, $n = 12$ and $m = 6$, where $\sigma$ is the (finite) distance at which the inter-particle potential is zero, and $\epsilon$ is the depth of the potential well.
However, we do not need to concern these exact values as what we want is the mathematical properties of the L-J potential.
}
The positive and negative terms correspond to the repulsive and attractive interactions respectively.
Note that this general form subsumes both the Newtonian gravitational potential and electrostatic potential.
The L-J potential is also applied in other fields like
the 
AI of chasing and evading in computer games \cite{LJAI}.

Recall that we consider the repulsive effect of the packets belonging to the same batch because these packets are vulnerable against burst errors when they are close to each other.
When we only consider repulsive interactions,\footnote{%
	We can use the attractive interactions to tune the packet separations so that they are not too far away, as we know that the benefit of using a larger separation is diminishing.
	However, a block is not very large in our application, so we do not consider this application in this paper.
}
the above potential energy models has a common form $A r^{-n}$ for some $A \ge 0$ and $n > 0$. %
This form is a strictly decreasing convex function with respect to $r$.
To reduce the computational cost, we set $A = 1$,
which has the following interpretations:
\begin{enumerate}
	\item 
		assign a unit quantity for all the packets, e.g., a unit mass $m_1 = m_2 = 1$ and a unit charge $q_1 = q_2 = 1$ in Newtonian gravitational potential and electrostatic potential respectively, so
		there is no packet being more repulsive than the others;
	\item 
		cancel out the constant involved in the model, e.g., the gravitational constant and the Coulomb's constant in Newtonian gravitational potential and electrostatic potential respectively; and
	\item 
		absorb the unit of $r$ so that we can use units other than the SI unit meter.
\end{enumerate}
We define a family of potential energies $\pe_n \colon \mathbb{R}^+ \to \mathbb{R}$ %
for $n > 0$
by \footnote{%
	As we use the potential energy to measure the goodness of the distance between two packets in a transmission sequence, it does not make sense to consider the distance between a packet and itself.
	Therefore, we do not define $\pe_n(r)$ for $r = 0$.
}
\begin{equation*}
	\pe_n(r) := r^{-n}, \quad r \neq 0.
\end{equation*}
Notice that we have $n = 1$ in both Newtonian gravitational potential and electrostatic potential.
So, $n = 1$ is a natural choice which can at the same time reduces the time for computation.
We will show in Section~\ref{sec:tp} that we can achieve a nice throughput with $n = 1$.

\subsection{Potential Energy Minimization}

In an intrablock interleaver, we are only concerned about the separation of the packets belonging to the same batch.
The packets in different batches do not interfere with each other and they can be sent consecutively.
To model this effect, we consider $L$ types of particles where
		the interactions between particles of different types are negligible, and
		the particles of the same type repulse each other.
The overall potential energy of the particles of the same type is the sum of the potential energies of each unordered pair of distinct particles of the same type.
If there is only one particle for a certain type, this single particle can be assigned to an arbitrary location, so we do not include this particle in the sum of potential energies. %
Recall that the smaller the potential energy the better the spread of the packets.
That is, we want to minimize the system energy, which is the sum of the overall potential energies of different types of particles.
Note that the interaction of index $i$ acting on index $j$ has the same magnitude as the interaction of index $j$ acting on index $i$.
Due to this symmetry, we can instead minimize half of the system energy to further reduce computation. %
We now formulate the potential energy minimization problem.
The variables $x_{k,i}$ are those denoted in \eqref{eq:s^-1} which assemble a transmission sequence.
As we can assign only one packet to each location in the sequence, we have %
$x_{k,i} \neq x_{k',i'}$ for all $(k,i) \neq (k',i')$. %
By symmetry breaking, we can reduce the search space by introducing a constraint $x_{k,i} < x_{k,i+1}$ for all $i \in \{1, 2, \ldots, t_k-1\}$.
If we consider all-pair interactions, then our problem is modeled as
\begin{equation}
\tag{PE} \label{eq:PE}
\begin{IEEEeqnarraybox}[][c]{rCl}
	\min_{x_{k,i} \in \mathcal{M}, \forall k, i} & \quad & \sum_{k = 1}^L \sum_{i = 1}^{t_k-1} \sum_{j = i+1}^{t_k} \pe_n(x_{k,j}-x_{k,i})\\
	\mathrm{s.t.} && x_{k,i} \neq x_{k',i'}, \forall (k,i) \neq (k',i'),\\ %
				  && x_{k,i} < x_{k,i+1}, \forall k, i.
\end{IEEEeqnarraybox}
\end{equation}
Note that we omitted the absolute value in expressing the separation $|x_{k,j}-x_{k,i}|$ as we have $j > i$.
On the other hand, if we consider neighboring interactions only as in the Ising model, then our problem is modeled as
\begin{equation}
\tag{APE} \label{eq:U}
\begin{IEEEeqnarraybox}[][c]{rCl}
	\min_{x_{k,i} \in \mathcal{M}, \forall k, i} & \quad & \sum_{k = 1}^L \sum_{i = 1}^{t_k-1} \pe_n(x_{k,i+1}-x_{k,i})\\
	\mathrm{s.t.} && x_{k,i} \neq x_{k',i'}, \forall (k,i) \neq (k',i'),\\ %
				  && x_{k,i} < x_{k,i+1}, \forall k, i.
\end{IEEEeqnarraybox}
\end{equation}

In \eqref{eq:U}, we only have to sum up $\mathcal{O}(t_k)$ terms instead of $\mathcal{O}(t_k^2)$ terms as in \eqref{eq:PE} for each batch $b_k$.
The constraint $x_{k,i} < x_{k,i+1}$ for all $i \in \{1, 2, \ldots, t_k-1\}$ ensures that the model only consider consecutive packets belonging to the same batch.

\subsection{Interpretation in Economics}

In economics, a \emph{utility} represents the preference ordering of a consumer over a choice set \cite{utility}.
The concept of utility is widely adopted in different areas such as risk management \cite{risk_utility}, game theory \cite{game_utility} and computer network \cite{lin2006tutorial,NUM}.

\emph{Utility maximization problem} is an optimization problem which maximizes the utility subject to certain constraints \cite{microeconomics}.
The decentralization problem describes the situation when each party maximizes its own utility, where the social planner problem is to maximize the sum of the utilities of all the parties \cite{social_planner}.
To illustrate the concepts, Hotelling's model \cite{hotelling} shows that, when there are two ice-cream sellers on the beach, the decentralized outcome would be both of them position themselves right at the middle of the beach, so each seller would get half of the consumers.
For the social planner, or the socially optimal outcome, it would be the best if they place themselves at the first quartile and the third quartile of the beach, the customer from the ends would not need to walk so far to reach the sellers, while both sellers still take half of the customers. %
The social planner problem is related to the problem investigated in this paper.

The utility in the setting of our problem can be regarded as a measure of satisfaction of a customer on the amount of resource received.
A customer receiving more resource would be more satisfied, but the increment would be smaller when the customer has already held a certain amount of resource.
One example would be the income utility, one dollar will mean more to the poor than to the rich, so-called the decreasing marginal utility of income \cite{microeconomics}.
Therefore, a utility $U(x)$ is usually defined as an increasing and concave function with respect to the amount of resource $x$ received.
In some contexts, there may be more assumptions on the utility, e.g., completely monotone utility
\cite{neg_sd}.
Some commonly used utilities in the performance analysis of computer networks are $\log$ and $\tan^{-1}$, which correspond to the congestion control algorithms of TCP Vegas and TCP Reno respectively \cite{low2017analytical}.

The idea of finding a transmission sequence via utility maximization is that we model the packet separation between two consecutive packets belonging to the same batch as the resources.
Then, a larger separation results in a higher utility, but the increment is slower when the separation is larger.
The overall utility is therefore the sum of the utilities of each separation involved.
This way, we can formulate problems similar to \eqref{eq:PE} and \eqref{eq:U} by replacing $\min$ by $\max$ and $\pe_n(\cdot)$ by $U(\cdot)$.
In other words, we can consider $-\pe_n$ as a representation of utility.

\section{Optimizing the Intrablock Interleaving Model}
\label{sec:solver}

Note that \eqref{eq:DE} is still a combinatorial optimization problem but the size of the search space $|\mathcal{F}(\{t_k\}_{k = 1}^L)| = \binom{T}{t_1, t_2, \ldots, t_L}$ is much smaller than that of \eqref{eq:BAR}.
In practice, we need to decide a transmission sequence quickly or otherwise the node would induce a huge delay.
So, instead of finding an exact optimum, we construct a ``good enough'' transmission sequence as an efficient approximation. %

In this section, we first propose a general transmission sequence approximation scheme which is not specific to any dispersion efficiency.
Then, we propose an optional fine-tuning scheme which tunes the transmission sequence into the one having a higher dispersion efficiency.
To see that our schemes gives a close-to-optimal transmission sequence, we compare them with the transmission sequence given by simulated annealing \cite{sa} and the true optimal transmission sequence given by constraint programming \cite{cp} in Section~\ref{sec:sim_de}.

\subsection{General Transmission Sequence Approximation}
\label{sec:alg:init}

If two consecutive packets of a batch are put close to each other in the transmission sequence, then they are vulnerable against burst errors.
To be fair, we want the packets of the same batch have a similar resistance to burst errors.
That is, we aim to separate the packets of the same batch uniformly.
Also, we also want to have a larger separation between the packets in order to increase its resistance to burst errors.

Note that we can only send $T$ packets in each transmission sequence.
That is, the best separation for a batch is to occupy the very beginning and the very last positions of the transmission sequence.
Even if we occupy these two positions, a batch which has a larger number of recoded packets has a smaller separation than that of a batch which has a smaller number of recoded packets.
In other words, in our best effort, the packets of the former batch are still more vulnerable against burst errors than the latter batch.
Due to this reason, we should give the two ends of a transmission sequence to the batch which has the largest number of recoded packets.

The same idea follows after we assign all the packets of such batch into the transmission sequence.
As the two ends of the transmission sequence are occupied, we would let the batch which has the second largest number of recoded packets to occupy the leftmost and rightmost unassigned positions, and so on and so forth.

When we consider a uniform separation, the time we want to transmit a packet may not be an integer unit of time.
It is also possible that the position in the transmission sequence has been occupied by another batch already.
To resolve these issues, we introduce a \emph{slip function} to ``slip'' a time to the closest unassigned position in the transmission sequence.

Before we define the slip function, we first define some notations for simplicity.
The very first position of the transmission sequence is called position $1$, the next position is called position $2$, and so on and so forth.
Position $1$ is the leftmost position.
To introduce the concept of the closest position around a time $x$, we define 
\begin{IEEEeqnarray*}{rCl}
	\underline{x} & := & \begin{cases}
		\max\{k \le \lfloor x \rfloor \colon \text{position } k \text{ is unassigned}\} & \text{if exists},\\
		0 & \text{otherwise};
	\end{cases}\\
	\overline{x} & := & \begin{cases}
		\min\{k \ge \lceil x \rceil \colon \text{position } k \text{ is unassigned}\} & \text{if exists,}\\
		0 & \text{otherwise}.
	\end{cases}
\end{IEEEeqnarray*}
Simply speaking, $\underline{x}$ is the closest unassigned position from the left, and $\overline{x}$ is the closest unassigned position from the right.
As the leftmost position is position $1$, a zero in the above definitions indicates that there is no unassigned position from the correspond side.
Note that we have enough positions to assign all the packets, so it is guaranteed that at least one of $\underline{x}$ or $\overline{x}$ is non-zero.
Now, we formally define the slip function by
\begin{equation*}
	\slip(x) := \begin{cases}
		\underline{x} & \text{if } [\overline{x} = 0] \vee [(\underline{x} \neq 0) \wedge (x - \underline{x} \le \overline{x} - x)],\\
		\overline{x} & \text{otherwise}.
	\end{cases}
\end{equation*}
The condition for $\slip(x) = \underline{x}$ can be interpreted this way.
First, if there is no unassigned position from the right, the closest unassigned position must be from the left.
Next, if there are unassigned positions from both sides, we compare the distance between $x$ and both $\underline{x}$ and $\overline{x}$.
If $\underline{x}$ is closer to $x$ or in the case of break-even, we slip to the left.

We need special handling if more than one batch has the same number of recoded packets.
As a simple example, suppose all batches have the same number of recoded packets, which is a case where a block interleaver works well.
Suppose the first batch occupies the two ends of the transmission sequence so that this batch has the largest separation.
The second batch will then have a relatively smaller separation than the first batch as the distance between the leftmost and rightmost unassigned positions becomes shorter.
This differ the burst resistance ability of these batches although they have the same number of recoded packets.
In other words, in this example, if the first batch occupies position $1$, then it should leave position $T$ to another batch.

In a general sense, we group the batches having the same number of recoded packets into a \emph{bundle}.
Let position $\ell$ and position $r$ be the leftmost and rightmost unassigned positions respectively.
Let $t > 1$ be the number of recoded packets of each of these batches.
Say, if the bundle has $b$ batches in it, ideally, these batches should occupy the leftmost $b$ positions $\ell, \ell+1, \ldots, \ell+b-1$ and the rightmost $b$ positions $r-b+1, r-b+2, \ldots, r$.
That is, for the batch in the bundle which occupies position $\ell+i$ for some $i$, it would occupy position $r-b+1+i$ on the other side.
Therefore, the best separation of each batch in the bundle is $((r-\ell+1)-b)/(t-1)$, which is also valid when $b = 1$.
After that, we find all the positions the first packet of each batch in this bundle would occupy, and then assign them to the batches in a round-robin manner, and so on and so forth.
We will explain the reason of doing this way by an example later in this subsection.

Lastly, if there is only one recoded packet in a batch, we do not have the concept of packet separation for this batch.
That is, we can put this packet at any position in the transmission sequence.

\begin{figure}
\centering
\removelatexerror
\begin{algorithm}[H]
	\small
	\caption{Transmission Sequence Approximation}
	\label{alg:init}
	\KwData{$\{t_i\}_{i = 1}^L$ where $t_1 \ge t_2 \ge \ldots \ge t_L > 0$, $\sum_{i = 1}^L t_i = T$.}
	\KwResult{A transmission sequence.}
	$\ell \leftarrow 1$; $r \leftarrow T$; $i \leftarrow 1$\;
	$f \leftarrow$ a $T$-integers array where index starts from $1$\;
	\While{$i \le  L$}{
		\If{$t_i > 1$}{ \nllabel{alg:init:ti>1}
			$\mathtt{bundle} \leftarrow \max \{j \colon t_j = t_i\} - i + 1$\;
			$\mathtt{gap} \leftarrow ((r-\ell+1) - \mathtt{bundle})/(t_i-1)$\;
			\For{$t = 1, 2, \ldots, t_i$}{ \nllabel{alg:init:ti>1:t}
				$\mathtt{pos} \leftarrow \emptyset$\;
				\For{$j = i, i+1, \ldots, i+\mathtt{bundle}-1$}{
					$\mathtt{tmp} \leftarrow \slip(\ell+(j-i)+(t-1)\mathtt{gap})$\;
					$\mathtt{pos} \leftarrow \mathtt{pos} \cup \{\mathtt{tmp}\}$\;
					Mark $f[\mathtt{tmp}]$ as assigned\; \nllabel{alg:init:ti>1:mark}
				}
				\For{$j = i, i+1, \ldots, i+\mathtt{bundle}-1$}{
					$f[\min \mathtt{pos}] \leftarrow j$\;
					$\mathtt{pos} \leftarrow \mathtt{pos} \setminus \{\min \mathtt{pos}\}$\;
				}
			} \nllabel{alg:init:ti>1:t:end}
			$i \leftarrow i + \mathtt{bundle}$\; \nllabel{alg:init:ti>1:end}
		}\Else{
			$f[\ell] \leftarrow i$; $i \leftarrow i+1$\; \nllabel{alg:init:ti=1}
		}
		$\ell \leftarrow$ the leftmost unassigned index of $f$\;
		$r \leftarrow$ the rightmost unassigned index of $f$\;
	}
	\Return $f$\;
\end{algorithm}
\end{figure}

Algorithm~\ref{alg:init} presents the pseudocode of the above discussion.
The transmission sequence constructed by this algorithm can be regarded as a good approximation to an optimal transmission sequence, where we will verify this in Section~\ref{sec:sim_de}.
This sequence can be used as an initial state for other methods to produce a further improved sequence.
In the algorithm, the transmission sequence, denoted by $f$, is represented by an array of $T$ integers.
Without loss of generality, assume $t_1 \ge t_2 \ge \ldots \ge t_L$.
Also, we can simply remove batches which have $0$ recoded packet and reduce the value of $L$ in the algorithm so that we can assume $t_L > 0$.
The variable $i$ corresponds to the $i$-th batch.
If $t_i > 1$, we find the size of the bundle and calculate the ``gap'', which is the best separation of each batch in the bundle.
Then, we assign the packets for the bundle as discussed above.
If $t_i = 1$, we assign the only packet of the batch to the leftmost unassigned position.

The following is an example to show the partially constructed transmission sequence after each iteration.
Consider $(t_1, t_2, \ldots, t_8) = (6, 5, 4, 3, 3, 2, 2, 2)$.
The visible spaces \text{\textvisiblespace} represent unassigned positions.
\begin{IEEEeqnarray*}{LCC}
	i = 1 & \qquad &
	\mathtt{1\text{\textvisiblespace\textvisiblespace\textvisiblespace\textvisiblespace}1\text{\textvisiblespace\textvisiblespace\textvisiblespace\textvisiblespace}1\text{\textvisiblespace\textvisiblespace\textvisiblespace\textvisiblespace\textvisiblespace}1\text{\textvisiblespace\textvisiblespace\textvisiblespace\textvisiblespace}1\text{\textvisiblespace\textvisiblespace\textvisiblespace\textvisiblespace}1}\\
	i = 2 &&
	\mathtt{12\text{\textvisiblespace\textvisiblespace\textvisiblespace}1\text{\textvisiblespace}2\text{\textvisiblespace\textvisiblespace}1\text{\textvisiblespace\textvisiblespace}2\text{\textvisiblespace\textvisiblespace}1\text{\textvisiblespace\textvisiblespace}2\text{\textvisiblespace}1\text{\textvisiblespace\textvisiblespace\textvisiblespace}21}\\
	i = 3 &&
	\mathtt{123\text{\textvisiblespace\textvisiblespace}1\text{\textvisiblespace}2\text{\textvisiblespace}31\text{\textvisiblespace\textvisiblespace}2\text{\textvisiblespace\textvisiblespace}13\text{\textvisiblespace}2\text{\textvisiblespace}1\text{\textvisiblespace\textvisiblespace}321}\\
	i = 4, 5 \text{ (bundle)} &&
	\mathtt{123451\text{\textvisiblespace}2\text{\textvisiblespace}31\text{\textvisiblespace}425\text{\textvisiblespace}13\text{\textvisiblespace}2\text{\textvisiblespace}145321}
\end{IEEEeqnarray*}
Right now, we explain why we do not assign $j$ to $f[\mathtt{tmp}]$ in Line~\ref{alg:init:ti>1:mark} directly.
If we follow Algorithm~\ref{alg:init} to fill in the last bundle, we will obtain
\begin{equation} \label{eq:seq_gen}
	\mathtt{123451\textcolor{red}{6}2\textcolor{blue}{7}31\textcolor{ForestGreen}{8}425\textcolor{red}{6}13\textcolor{blue}{7}2\textcolor{ForestGreen}{8}145321}
\end{equation}
where the separations of the packets in the $6$-th, the $7$-th and the $8$-th batches are $9$, $10$ and $9$ respectively.
However, if we assign $j$ to $f[\mathtt{tmp}]$ directly, we will obtain
\begin{equation*}
	\mathtt{123451\textcolor{red}{6}2\textcolor{blue}{7}31\textcolor{ForestGreen}{8}425\textcolor{ForestGreen}{8}13\textcolor{red}{6}2\textcolor{blue}{7}145321}
\end{equation*}
where the separations of the packets in the last three batches are $12$, $12$ and $4$ respectively, which makes the packets in the last batch much more vulnerable against burst errors.

\subsection{Fine-Tuning}

The transmission sequence produced by Algorithm~\ref{alg:init} does not depend on a specific dispersion efficiency.
It is possible that we can further manipulate the sequence into a better one in the sense of having a higher dispersion efficiency.

In this subsection, we introduce a simple fine-tuning algorithm which try to search for a better sequence by swapping adjacent packets.
We take the transmission sequence \eqref{eq:seq_gen} produced by Algorithm~\ref{alg:init} as an example.
The separations of the consecutive packets of the $1$-st batch and that of the $6$-th batch are illustrated below.
\begin{equation*}
	\mathtt{\rlap{$\overbrace{\phantom{123451}}^5$}12345\rlap{$\overbrace{\phantom{162731}}^5$}1\rlap{$\underbrace{\phantom{6273184256}}_9$}6273\rlap{$\overbrace{\phantom{1842561}}^6$}18425\textcolor{red}{6}\rlap{$\overbrace{\phantom{137281}}^5$}\textcolor{blue}{1}3728\rlap{$\overbrace{\phantom{145321}}^5$}145321}
\end{equation*}
We can see that if we swap the colored $1$ and $6$, the set of separations for the $1$-th batch remains unchanged, but the separation for the $6$-th batch is increased by $1$.
That is, we have increased the resistance to burst errors of the $6$-th batch a little bit without changing the one for the $1$-st batch.

\begin{figure}
\removelatexerror
\begin{algorithm}[H]
	\small
	\caption{Fine-Tuning Adjacent Packets}
	\label{alg:tune}
	\KwData{A transmission sequence $f$.}
	\KwResult{A fine-tuned transmission sequence.}
	\For{$i = 1, 2, \ldots, T-1$}{
		\If{swapping $f[i]$ and $f[i+1]$ gives larger $\eff(f)$}{
			Swap $f[i]$ and $f[i+1]$\;
			Restart the loop\;
		}
	}
	\Return $f$ \;
\end{algorithm}
\end{figure}

Algorithm~\ref{alg:tune} is our simple fine-tuning algorithm.
We scan through the transmission sequence once and try whether we can improve the dispersion efficiency by swapping adjacent packets.
If we can find one, then we swap the packets and scan from the beginning again to see whether we can further improve the sequence.
As the number of packets in a block, $T$, is not large in practice, this fine-tuning algorithm can still be run efficient.
We remark that this fine-tuning step is optional as we will show in the next section that the approximation given by Algorithm~\ref{alg:init} is already good enough.

\section{Numerical Evaluations}
\label{sec:sim}

\subsection{Packet Loss Models}
\label{sec:lossmodel}
As we solve the adaptive recoding problem in a block-by-block manner, the channel model and parameters in each block is independent of that of another block.
In other words, if the network has feedback available, we can update the channel model and parameters for the upcoming blocks.
As a block is not very large, we can assume that the channel condition remains the same within the block.

To apply adaptive recoding in practice, we need to describe the channel by certain mathematical model although we do not assume a specific one in our design.
Most commonly used channel models such as those listed below are stationary stochastic processes, thus their packet loss pattern are also a stationary stochastic processes \cite{protocol}.
Some candidates of channel models are:

\subsubsection{Independent Packet Loss Model}
The burst loss has been alleviated by interleaving so that the loss pattern is ``closer'' to independent loss. This independent loss model is used in various works in BNC including \cite{zhou17b,variable,rf}.

\subsubsection{Gilbert-Elliott (GE) Model}
A two-state hidden Markov model \cite{GilbertBurst,ElliottBurst} for bursty channel which is widely-used in the literature \cite{hasslinger08,gb_eg1,gb_eg2} and also for multi-hop networks \cite{ge_adaptive,frohn11,bittt_space}. The parameters can be trained by the Baum-Welch algorithm \cite{hmm}.

\subsubsection{Multiple-State Markov Chain Model}
Model the burst loss pattern by Markov chain having more than two states \cite{nstate,morestate}, which can capture physical properties like BPSK coding with Rayleigh fading process for wireless channels \cite{wang95,ge_pkt}.

\begin{figure}
	\centering
	\begin{tikzpicture}[scale=1,font=\footnotesize,state/.style={circle,draw=black,thick,inner sep=1pt,minimum size=20pt,align=center},->,>=stealth',thick,every node/.style={transform shape}]
		\node[state] (G) at (0,0) {\textbf{G}\\$g$};
		\node[state] (B) at (3,0) {\textbf{B}\\$b$};
		\path (G) edge [bend left] node[below] {$p$} (B);
		\path (B) edge [bend left] node[above] {$q$} (G);
		\path (G) edge [loop left] node[left] {$1-p$} (G);
		\path (B) edge [loop right] node[right] {$1-q$} (B);
	\end{tikzpicture}
	\caption{The Gilbert-Elliott model for burst packet loss pattern.
	}
	\label{fig:ge_model}
\end{figure}
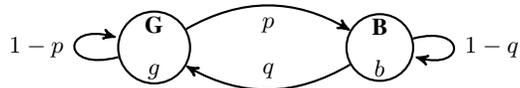

For the sake of demonstration, we apply a GE model at each link as an example to evaluate some functions and simulations in this paper.
The mathematics involved can be found in Appendix~\ref{sec:ge}.
A GE model, as illustrated in Fig.~\ref{fig:ge_model}, consists of two states, \textbf{G} and \textbf{B}, which are called the good state and bad state respectively.
In each state, there is an independent event to decide whether a packet is lost or not.
The probabilities of losing a packet in \textbf{G} and \textbf{B} are $g$ and $b$ respectively.
As a reasonable channel model, we assume the transition probabilities $0 < p, q < 1$ so that the Markov chain is ergodic.
Note that the independent packet loss model can be subsumed under the GE model, e.g., $p = 1-q$, $g = 0$ and $b = 1$.
When a node transmits a packet, the state of the Markov chain of the GE model is changed according to its transition matrix and the outcome indicates whether the packet is received or not by the next node.
Each change of states is actually describing the stochastic process under the observation of a unit time.
In case the timescale of a model and the one for packet transmissions are not synchronized, there is another model which describe the observation with a synchronized timescale.
For example, if the Markov chain changes the states twice when the node transmits a packet, then the channel model is equivalent to a Markov chain where its transition matrix is the square of that of the former chain.
We can also produce a transition matrix for a short-term observation by taking the root of the transition matrix for a longer observation, which is a technique applied in the models for credit rating in finance \cite{credit} and the progression of chronic disease in medicine \cite{transition}.

In general, we can normalize the timescale of a Markov chain by taking a non-negative power of its transition matrix, but beware that the non-negative power of an arbitrary stochastic matrix may not be stochastic \cite{pth}.
Luckily, the non-negative power of the transition matrix of a GE model is stochastic.
See Appendix~\ref{sec:ge:pid} for the verification.
In our numerical evaluations, we only consider GE models where their timescales and that for packet transmissions are synchronized.

\subsection{Settings}

We use the same network topology as in \cite{ge_adaptive}, which uses a GE model for BNC, to evaluate the performance of our intrablock interleaver.
We consider an $8$-hop line network where each link uses a GE model with the same set of parameters.

\begin{table}
	\centering
	\scriptsize
	\caption{The average packet loss rate $\varepsilon$ and average burst error length $\mathsf{ABEL}$ of the Gilbert-Elliott Models where $g = 0.1$ and $b = 0.8$}
	\label{tab:channels}
	\begin{tabular}{c|cc|cc|cc}
		\multirow{2}{*}{\diagbox{$\mathsf{ABEL}$}{$\varepsilon$}} & \multicolumn{2}{c|}{$35\%$} & \multicolumn{2}{c|}{$45\%$} & \multicolumn{2}{c}{$55\%$}\\ \cline{2-7}
		& $p$ & $q$ & $p$ & $q$ & $p$ & $q$\\
		\toprule
		$2$ & $4/21$ & $ 12/35$ & $20/49$ & $20/49$ & $4/5$ & $4/9$\\
		$2.5$ & $5/63$ & $1/7$ & $11/49$ & $11/49$ & $17/35$ & $17/63$\\
		$900/299$ & $23/5670$ & $23/3150$ & $1/10$ & $1/10$ & $859/3150$ & $859/5670$\\
		\bottomrule
	\end{tabular}
\end{table}

The parameters of the GE model used in \cite{ge_adaptive} are $p = 0.1$, $q = 0.1$, $g = 0.1$ and $b = 0.8$.
The corresponding average packet loss rate $\varepsilon$ and average burst error length $\mathsf{ABEL}$ of this set of parameters are $45\%$ and $900/299 \approx 3.01$ respectively (see Appendix~\ref{sec:ge:burst} for the method to calculate $\varepsilon$ and $\mathsf{ABEL}$ from the parameters, and vice versa).
We consider the set of parameters as one of the cases in our evaluations.
For a fair comparison, we also evaluate other sets of parameters with the same $g = 0.1$ and $b = 0.8$ but with different $\varepsilon$ and $\mathsf{ABEL}$.
Table~\ref{tab:channels} lists the parameters of the GE models we used in our numerical evaluations.

We evaluate batch sizes $M = 4, 8$ and block sizes $L = 4, 8$.
For each pair of batch size $M$ and block size $L$, we fix the number of packets in each block to be $T = ML$.
We use $-\pe_1(x) := -\frac{1}{x}$, $-\pe_2(x) := -\frac{1}{x^2}$, $\ln(x)$ and $\tan^{-1}(x)$ to evaluate the dispersion efficiencies.

\subsection{Benchmarks}

The fine-tuned transmission sequence given by Algorithm~\ref{alg:tune} may not be optimal.
We try to search for a better transmission sequence in another local optimum by simulated annealing (SA) \cite{sa}.
Note that SA is a metaheuristic approach consuming a much higher computational power so that it may not be suitable to be used to generate transmission sequence in real deployment.
Nevertheless, we can use SA for comparison purpose in simulations.
On the other hand, we can find the true optimal dispersion efficiency by constraint programming (CP) \cite{cp}.
However, each transmission sequence takes a long time to be generated so we only apply CP on some samples to verify the quality of the transmission sequence generated by our algorithms in terms of dispersion efficiencies.
The detailed formulations of our problem in SA and CP can be found in Appendix~\ref{sec:sa} and \ref{sec:cp} respectively.

To compare with the throughput of the intrablock interleaver, two basic benchmarks are the throughput of baseline recoding with a block interleaver and the one given by the batch-stream interleaver with adaptive recoding.
The throughput given by the intrablock interleaver is expected to lay between the two above benchmarks.

\subsection{Dispersion Efficiency}
\label{sec:sim_de}

\begin{table}
	\centering
	\scriptsize
	\caption{Comparison of Dispersion Efficiencies}
	\label{tab:de}
	\begin{tabular}{l|c|ccc|cc}
		& CP & Alg.~\ref{alg:init} & Alg.~\ref{alg:tune} & SA & worst & SA worst \\
		\toprule
		\multicolumn{7}{c}{$\mathbf{t} = (6, 5, 4, 3, 3, 2, 2, 2)$, from the example in Section~\ref{sec:alg:init}}\\
		$\effpe_{-\pe_1}$ &     -4.167 & -4.178 & -4.167 & -4.167 & -27.450 & -4.667 \\
		$\effpe_{-\pe_2}$ &     -0.531 & -0.534 & -0.531 & -0.531 & -22.644 & -0.768 \\
		$\effpe_{\ln}$ &        95.028 & 94.940 & 95.028 & 95.028 & 19.985 & 93.100  \\
		$\effpe_{\tan^{-1}}$ &  58.691 & 58.679 & 58.691 & 58.691 & 39.946 & 58.068  \\
		\midrule
		$\effu_{-\pe_1}$ &      -2.756 & -2.788 & -2.757 & -2.757 & -19.000 & -2.991 \\
		$\effu_{-\pe_2}$ &      -0.428 & -0.435 & -0.428 & -0.428 & -19.000 & -0.573 \\
		$\effu_{\ln}$ &         37.721 & 37.064 & 37.711 & 37.711 & 0.000 & 36.933   \\
		$\effu_{\tan^{-1}}$ &   27.114 & 27.080 & 27.111 & 27.111 & 14.923 & 26.606  \\
		\specialrule{.7pt}{1pt}{1pt}

		\multicolumn{7}{c}{$\mathbf{t} = (5, 3, 3, 5)$, a sample when $L = 4$, $M = 4$}\\
		$\effpe_{-\pe_1}$ &     -4.814 & -4.857 & -4.814 & -4.814 & -17.833 & -4.843 \\
		$\effpe_{-\pe_2}$ &     -1.080 & -1.105 & -1.082 & -1.082 & -14.569 & -1.204 \\
		$\effpe_{\ln}$ &        46.757 & 46.627 & 46.757 & 46.757 & 12.712 & 46.530  \\
		$\effpe_{\tan^{-1}}$ &  36.115 & 36.075 & 36.115 & 36.115 & 25.930 & 36.033  \\
		\midrule
		$\effu_{-\pe_1}$ &      -3.167 & -3.250 & -3.167 & -3.167 & -12.000 & -3.283 \\
		$\effu_{-\pe_2}$ &      -0.875 & -0.910 & -0.875 & -0.875 & -12.000 & -0.952 \\
		$\effu_{\ln}$ &         16.296 & 15.890 & 16.296 & 16.296 & 0.000 & 15.956   \\
		$\effu_{\tan^{-1}}$ &   15.762 & 15.683 & 15.762 & 15.762 & 9.425 & 15.716   \\
		\specialrule{.7pt}{1pt}{1pt}

		\multicolumn{7}{c}{$\mathbf{t} = (3, 5, 8, 0)$, a sample when $L = 4$, $M = 4$}\\
		$\effpe_{-\pe_1}$ &     -9.136 & -9.136 & -9.136 & -9.136 & -22.660 & -9.168\\
		$\effpe_{-\pe_2}$ &     -2.868 & -2.868 & -2.868 & -2.868 & -16.786 & -3.009\\
		$\effpe_{\ln}$ &        69.327 & 69.327 & 69.327 & 69.327 & 31.911 & 68.761\\
		$\effpe_{\tan^{-1}}$ &  55.603 & 55.603 & 55.603 & 55.603 & 45.014 & 55.603\\
		\midrule
		$\effu_{-\pe_1}$ &      -5.083 & -5.117 & -5.117 & -5.117 & -13.000 & -5.408\\
		$\effu_{-\pe_2}$ &      -2.187 & -2.200 & -2.200 & -2.200 & -13.000 & -2.453\\
		$\effu_{\ln}$ &         13.000 & 12.818 & 12.818 & 12.818 & 0.000 & 12.984\\
		$\effu_{\tan^{-1}}$ &   15.631 & 15.599 & 15.599 & 15.631 & 10.210 & 15.551\\
		\specialrule{.7pt}{1pt}{1pt}

		\multicolumn{7}{c}{$\mathbf{t} = (9, 8, 7, 8)$, a sample when $L = 4$, $M = 8$}\\
		$\effpe_{-\pe_1}$ &     -13.836 & -13.838 & -13.838 & -13.838 & -55.096 & -15.043\\
		$\effpe_{-\pe_2}$ &     -2.408  & -2.408 & -2.408 & -2.408 & -38.022 & -2.991\\
		$\effpe_{\ln}$ &        260.458 & 260.458 & 260.458 & 260.458 & 104.299 & 256.439\\
		$\effpe_{\tan^{-1}}$ &  163.829 & 163.829 & 163.829 & 163.829 & 129.645 & 163.009\\
		\midrule
		$\effu_{-\pe_1}$ &      -7.033  & -7.033 & -7.033 & -7.033 & -28.000 & -8.026\\
		$\effu_{-\pe_2}$ &      -1.776  & -1.776 & -1.776 & -1.776 & -28.000 & -2.251\\
		$\effu_{\ln}$ &         38.752  & 38.752 & 38.752 & 38.752 & -0.000 & 36.763\\
		$\effu_{\tan^{-1}}$ &   37.094  & 37.094 & 37.094 & 37.094 & 21.991 & 36.457\\
		\bottomrule
	\end{tabular}
\end{table}

We first present the comparison in dispersion efficiencies.
Table~\ref{tab:de} shows the dispersion efficiencies of the transmission sequence given by different approaches corrected to $3$ decimal places.
For brevity, we write $\mathbf{t} = (t_1, t_2, \ldots, t_L)$ in the table.
Algorithm~\ref{alg:tune} takes the output of Algorithm~\ref{alg:init} as the initial transmission sequence.
The column ``CP'' is the optimal transmission sequence solved by CP solver.
The column ``SA'' is for SA which takes the output of Algorithm~\ref{alg:tune} as the initial state.
The column ``worst'' corresponds to the transmission sequence which has the worst dispersion efficiency, i.e., there is no interleaving so that the IPG of each pair of consecutive packets in each batch is always $1$.
The column ``SA worst'' is for SA with the worst transmission sequence as the initial state.

From the table, we can see that the fine-tuning approach can improve the dispersion efficiency from our approximation transmission sequence, which is very close to the true optimum.
SA can barely improve the dispersion efficiency from the fine-tuned transmission sequence.
We can also see that in most of the cases, SA can give a better transmission sequence when we use an initial state closer to some local optimum.

\subsection{Throughput}
\label{sec:tp}

\begin{figure}
	\centering
	\begin{tikzpicture}[scale=.73]
		\begin{groupplot}[group style={group size=3 by 3, ylabels at=edge left},
			small,
			grid=both,
			xtick={1,2,...,8},
			ytick={0.05,0.1,...,0.7},
			xlabel=Hops,
			ylabel=Normalized Throughput,
			label style={font=\footnotesize},
			x label style={at={(axis description cs:.5,.05)},anchor=north},
			y label style={at={(axis description cs:.12,.5)},anchor=south},
			xmin=1,xmax=8,
			cycle list name=color list,
			legend style={at={(.17,1.02)},anchor=south west,legend columns=6,font=\scriptsize}
		]
			\nextgroupplot[ymin=0.29,ymax=0.66]
			\node[anchor=north east] at (axis description cs:1,1) {$\varepsilon=35\%, \mathsf{ABEL}=2$};

			\addplot+[blue,mark=,dashed] table[x=N,y=B4,col sep=comma]{data/tpM4p1l1.txt};
			\addlegendentry{BR-BI, $L=4$};
			\addplot+[blue,mark=,densely dotted] table[x=N,y=A4,col sep=comma]{data/tpM4p1l1.txt};
			\addlegendentry{AR-SI, $L=4$};
			\addplot+[blue,mark=] table[x=N,y=I4,col sep=comma]{data/tpM4p1l1.txt};
			\addlegendentry{AR-IBI, $L=4$};

			\addplot+[red,mark=,dashed] table[x=N,y=B8,col sep=comma]{data/tpM4p1l1.txt};
			\addlegendentry{BR-BI, $L=8$};
			\addplot+[red,mark=,densely dotted] table[x=N,y=A8,col sep=comma]{data/tpM4p1l1.txt};
			\addlegendentry{AR-SI, $L=8$};
			\addplot+[red,mark=] table[x=N,y=I8,col sep=comma]{data/tpM4p1l1.txt};
			\addlegendentry{AR-IBI, $L=8$};

			\nextgroupplot[ymin=0.19,ymax=0.56]
			\node[anchor=north east] at (axis description cs:1,1) {$\varepsilon=45\%, \mathsf{ABEL}=2$};

			\addplot+[blue,mark=,dashed] table[x=N,y=B4,col sep=comma]{data/tpM4p2l1.txt};
			\addplot+[blue,mark=,densely dotted] table[x=N,y=A4,col sep=comma]{data/tpM4p2l1.txt};
			\addplot+[blue,mark=] table[x=N,y=I4,col sep=comma]{data/tpM4p2l1.txt};

			\addplot+[red,mark=,dashed] table[x=N,y=B8,col sep=comma]{data/tpM4p2l1.txt};
			\addplot+[red,mark=,densely dotted] table[x=N,y=A8,col sep=comma]{data/tpM4p2l1.txt};
			\addplot+[red,mark=] table[x=N,y=I8,col sep=comma]{data/tpM4p2l1.txt};

			\nextgroupplot[ymin=0.11,ymax=0.46]
			\node[anchor=north east] at (axis description cs:1,1) {$\varepsilon=55\%, \mathsf{ABEL}=2$};

			\addplot+[blue,mark=,dashed] table[x=N,y=B4,col sep=comma]{data/tpM4p3l1.txt};
			\addplot+[blue,mark=,densely dotted] table[x=N,y=A4,col sep=comma]{data/tpM4p3l1.txt};
			\addplot+[blue,mark=] table[x=N,y=I4,col sep=comma]{data/tpM4p3l1.txt};

			\addplot+[red,mark=,dashed] table[x=N,y=B8,col sep=comma]{data/tpM4p3l1.txt};
			\addplot+[red,mark=,densely dotted] table[x=N,y=A8,col sep=comma]{data/tpM4p3l1.txt};
			\addplot+[red,mark=] table[x=N,y=I8,col sep=comma]{data/tpM4p3l1.txt};

			\nextgroupplot[ymin=0.22,ymax=0.66]
			\node[anchor=north east] at (axis description cs:1,1) {$\varepsilon=35\%, \mathsf{ABEL}=2.5$};

			\addplot+[blue,mark=,dashed] table[x=N,y=B4,col sep=comma]{data/tpM4p1l2.txt};
			\addplot+[blue,mark=,densely dotted] table[x=N,y=A4,col sep=comma]{data/tpM4p1l2.txt};
			\addplot+[blue,mark=] table[x=N,y=I4,col sep=comma]{data/tpM4p1l2.txt};

			\addplot+[red,mark=,dashed] table[x=N,y=B8,col sep=comma]{data/tpM4p1l2.txt};
			\addplot+[red,mark=,densely dotted] table[x=N,y=A8,col sep=comma]{data/tpM4p1l2.txt};
			\addplot+[red,mark=] table[x=N,y=I8,col sep=comma]{data/tpM4p1l2.txt};

			\nextgroupplot[ymin=0.18,ymax=0.56]
			\node[anchor=north east] at (axis description cs:1,1) {$\varepsilon=45\%, \mathsf{ABEL}=2.5$};

			\addplot+[blue,mark=,dashed] table[x=N,y=B4,col sep=comma]{data/tpM4p2l2.txt};
			\addplot+[blue,mark=,densely dotted] table[x=N,y=A4,col sep=comma]{data/tpM4p2l2.txt};
			\addplot+[blue,mark=] table[x=N,y=I4,col sep=comma]{data/tpM4p2l2.txt};

			\addplot+[red,mark=,dashed] table[x=N,y=B8,col sep=comma]{data/tpM4p2l2.txt};
			\addplot+[red,mark=,densely dotted] table[x=N,y=A8,col sep=comma]{data/tpM4p2l2.txt};
			\addplot+[red,mark=] table[x=N,y=I8,col sep=comma]{data/tpM4p2l2.txt};

			\nextgroupplot[ymin=0.11,ymax=0.46]
			\node[anchor=north east] at (axis description cs:1,1) {$\varepsilon=55\%, \mathsf{ABEL}=2.5$};

			\addplot+[blue,mark=,dashed] table[x=N,y=B4,col sep=comma]{data/tpM4p3l2.txt};
			\addplot+[blue,mark=,densely dotted] table[x=N,y=A4,col sep=comma]{data/tpM4p3l2.txt};
			\addplot+[blue,mark=] table[x=N,y=I4,col sep=comma]{data/tpM4p3l2.txt};

			\addplot+[red,mark=,dashed] table[x=N,y=B8,col sep=comma]{data/tpM4p3l2.txt};
			\addplot+[red,mark=,densely dotted] table[x=N,y=A8,col sep=comma]{data/tpM4p3l2.txt};
			\addplot+[red,mark=] table[x=N,y=I8,col sep=comma]{data/tpM4p3l2.txt};

			\nextgroupplot[ymin=0.08,ymax=0.66]
			\node[anchor=north east] at (axis description cs:1,1) {$\varepsilon=35\%, \mathsf{ABEL}=\frac{900}{299}$};

			\addplot+[blue,mark=,dashed] table[x=N,y=B4,col sep=comma]{data/tpM4p1l3.txt};
			\addplot+[blue,mark=,densely dotted] table[x=N,y=A4,col sep=comma]{data/tpM4p1l3.txt};
			\addplot+[blue,mark=] table[x=N,y=I4,col sep=comma]{data/tpM4p1l3.txt};

			\addplot+[red,mark=,dashed] table[x=N,y=B8,col sep=comma]{data/tpM4p1l3.txt};
			\addplot+[red,mark=,densely dotted] table[x=N,y=A8,col sep=comma]{data/tpM4p1l3.txt};
			\addplot+[red,mark=] table[x=N,y=I8,col sep=comma]{data/tpM4p1l3.txt};

			\nextgroupplot[ymin=0.13,ymax=0.56]
			\node[anchor=north east] at (axis description cs:1,1) {$\varepsilon=45\%, \mathsf{ABEL}=\frac{900}{299}$};

			\addplot+[blue,mark=,dashed] table[x=N,y=B4,col sep=comma]{data/tpM4p2l3.txt};
			\addplot+[blue,mark=,densely dotted] table[x=N,y=A4,col sep=comma]{data/tpM4p2l3.txt};
			\addplot+[blue,mark=] table[x=N,y=I4,col sep=comma]{data/tpM4p2l3.txt};

			\addplot+[red,mark=,dashed] table[x=N,y=B8,col sep=comma]{data/tpM4p2l3.txt};
			\addplot+[red,mark=,densely dotted] table[x=N,y=A8,col sep=comma]{data/tpM4p2l3.txt};
			\addplot+[red,mark=] table[x=N,y=I8,col sep=comma]{data/tpM4p2l3.txt};

			\nextgroupplot[ymin=0.1,ymax=0.46]
			\node[anchor=north east] at (axis description cs:1,1) {$\varepsilon=55\%, \mathsf{ABEL}=\frac{900}{299}$};

			\addplot+[blue,mark=,dashed] table[x=N,y=B4,col sep=comma]{data/tpM4p3l3.txt};
			\addplot+[blue,mark=,densely dotted] table[x=N,y=A4,col sep=comma]{data/tpM4p3l3.txt};
			\addplot+[blue,mark=] table[x=N,y=I4,col sep=comma]{data/tpM4p3l3.txt};

			\addplot+[red,mark=,dashed] table[x=N,y=B8,col sep=comma]{data/tpM4p3l3.txt};
			\addplot+[red,mark=,densely dotted] table[x=N,y=A8,col sep=comma]{data/tpM4p3l3.txt};
			\addplot+[red,mark=] table[x=N,y=I8,col sep=comma]{data/tpM4p3l3.txt};

		\end{groupplot}
	\end{tikzpicture}
	\caption{The normalized throughput of BNC with $M = 4$.}
	\label{fig:tpM4}
\end{figure}

\begin{figure}
	\centering
	\begin{tikzpicture}[scale=.73]
		\begin{groupplot}[group style={group size=3 by 3, ylabels at=edge left},
			small,
			grid=both,
			xtick={1,2,...,8},
			ytick={0.05,0.1,...,0.7},
			xlabel=Hops,
			ylabel=Normalized Throughput,
			label style={font=\footnotesize},
			x label style={at={(axis description cs:.5,.05)},anchor=north},
			y label style={at={(axis description cs:.12,.5)},anchor=south},
			xmin=1,xmax=8,
			cycle list name=color list,
			legend style={at={(.17,1.02)},anchor=south west,legend columns=6,font=\scriptsize}
		]
			\nextgroupplot[ymin=0.39,ymax=0.66]
			\node[anchor=north east] at (axis description cs:1,1) {$\varepsilon=35\%, \mathsf{ABEL}=2$};

			\addplot+[blue,mark=,dashed] table[x=N,y=B4,col sep=comma]{data/tpM8p1l1.txt};
			\addlegendentry{BR-BI, $L=4$};
			\addplot+[blue,mark=,densely dotted] table[x=N,y=A4,col sep=comma]{data/tpM8p1l1.txt};
			\addlegendentry{AR-SI, $L=4$};
			\addplot+[blue,mark=] table[x=N,y=I4,col sep=comma]{data/tpM8p1l1.txt};
			\addlegendentry{AR-IBI, $L=4$};

			\addplot+[red,mark=,dashed] table[x=N,y=B8,col sep=comma]{data/tpM8p1l1.txt};
			\addlegendentry{BR-BI, $L=8$};
			\addplot+[red,mark=,densely dotted] table[x=N,y=A8,col sep=comma]{data/tpM8p1l1.txt};
			\addlegendentry{AR-SI, $L=8$};
			\addplot+[red,mark=] table[x=N,y=I8,col sep=comma]{data/tpM8p1l1.txt};
			\addlegendentry{AR-IBI, $L=8$};

			\nextgroupplot[ymin=0.29,ymax=0.56]
			\node[anchor=north east] at (axis description cs:1,1) {$\varepsilon=45\%, \mathsf{ABEL}=2$};

			\addplot+[blue,mark=,dashed] table[x=N,y=B4,col sep=comma]{data/tpM8p2l1.txt};
			\addplot+[blue,mark=,densely dotted] table[x=N,y=A4,col sep=comma]{data/tpM8p2l1.txt};
			\addplot+[blue,mark=] table[x=N,y=I4,col sep=comma]{data/tpM8p2l1.txt};

			\addplot+[red,mark=,dashed] table[x=N,y=B8,col sep=comma]{data/tpM8p2l1.txt};
			\addplot+[red,mark=,densely dotted] table[x=N,y=A8,col sep=comma]{data/tpM8p2l1.txt};
			\addplot+[red,mark=] table[x=N,y=I8,col sep=comma]{data/tpM8p2l1.txt};

			\nextgroupplot[ymin=0.2,ymax=0.46]
			\node[anchor=north east] at (axis description cs:1,1) {$\varepsilon=55\%, \mathsf{ABEL}=2$};

			\addplot+[blue,mark=,dashed] table[x=N,y=B4,col sep=comma]{data/tpM8p3l1.txt};
			\addplot+[blue,mark=,densely dotted] table[x=N,y=A4,col sep=comma]{data/tpM8p3l1.txt};
			\addplot+[blue,mark=] table[x=N,y=I4,col sep=comma]{data/tpM8p3l1.txt};

			\addplot+[red,mark=,dashed] table[x=N,y=B8,col sep=comma]{data/tpM8p3l1.txt};
			\addplot+[red,mark=,densely dotted] table[x=N,y=A8,col sep=comma]{data/tpM8p3l1.txt};
			\addplot+[red,mark=] table[x=N,y=I8,col sep=comma]{data/tpM8p3l1.txt};

			\nextgroupplot[ymin=0.34,ymax=0.66]
			\node[anchor=north east] at (axis description cs:1,1) {$\varepsilon=35\%, \mathsf{ABEL}=2.5$};

			\addplot+[blue,mark=,dashed] table[x=N,y=B4,col sep=comma]{data/tpM8p1l2.txt};
			\addplot+[blue,mark=,densely dotted] table[x=N,y=A4,col sep=comma]{data/tpM8p1l2.txt};
			\addplot+[blue,mark=] table[x=N,y=I4,col sep=comma]{data/tpM8p1l2.txt};

			\addplot+[red,mark=,dashed] table[x=N,y=B8,col sep=comma]{data/tpM8p1l2.txt};
			\addplot+[red,mark=,densely dotted] table[x=N,y=A8,col sep=comma]{data/tpM8p1l2.txt};
			\addplot+[red,mark=] table[x=N,y=I8,col sep=comma]{data/tpM8p1l2.txt};

			\nextgroupplot[ymin=0.28,ymax=0.56]
			\node[anchor=north east] at (axis description cs:1,1) {$\varepsilon=45\%, \mathsf{ABEL}=2.5$};

			\addplot+[blue,mark=,dashed] table[x=N,y=B4,col sep=comma]{data/tpM8p2l2.txt};
			\addplot+[blue,mark=,densely dotted] table[x=N,y=A4,col sep=comma]{data/tpM8p2l2.txt};
			\addplot+[blue,mark=] table[x=N,y=I4,col sep=comma]{data/tpM8p2l2.txt};

			\addplot+[red,mark=,dashed] table[x=N,y=B8,col sep=comma]{data/tpM8p2l2.txt};
			\addplot+[red,mark=,densely dotted] table[x=N,y=A8,col sep=comma]{data/tpM8p2l2.txt};
			\addplot+[red,mark=] table[x=N,y=I8,col sep=comma]{data/tpM8p2l2.txt};

			\nextgroupplot[ymin=0.2,ymax=0.46]
			\node[anchor=north east] at (axis description cs:1,1) {$\varepsilon=55\%, \mathsf{ABEL}=2.5$};

			\addplot+[blue,mark=,dashed] table[x=N,y=B4,col sep=comma]{data/tpM8p3l2.txt};
			\addplot+[blue,mark=,densely dotted] table[x=N,y=A4,col sep=comma]{data/tpM8p3l2.txt};
			\addplot+[blue,mark=] table[x=N,y=I4,col sep=comma]{data/tpM8p3l2.txt};

			\addplot+[red,mark=,dashed] table[x=N,y=B8,col sep=comma]{data/tpM8p3l2.txt};
			\addplot+[red,mark=,densely dotted] table[x=N,y=A8,col sep=comma]{data/tpM8p3l2.txt};
			\addplot+[red,mark=] table[x=N,y=I8,col sep=comma]{data/tpM8p3l2.txt};

			\nextgroupplot[ymin=0.13,ymax=0.66]
			\node[anchor=north east] at (axis description cs:1,1) {$\varepsilon=35\%, \mathsf{ABEL}=\frac{900}{299}$};

			\addplot+[blue,mark=,dashed] table[x=N,y=B4,col sep=comma]{data/tpM8p1l3.txt};
			\addplot+[blue,mark=,densely dotted] table[x=N,y=A4,col sep=comma]{data/tpM8p1l3.txt};
			\addplot+[blue,mark=] table[x=N,y=I4,col sep=comma]{data/tpM8p1l3.txt};

			\addplot+[red,mark=,dashed] table[x=N,y=B8,col sep=comma]{data/tpM8p1l3.txt};
			\addplot+[red,mark=,densely dotted] table[x=N,y=A8,col sep=comma]{data/tpM8p1l3.txt};
			\addplot+[red,mark=] table[x=N,y=I8,col sep=comma]{data/tpM8p1l3.txt};

			\nextgroupplot[ymin=0.23,ymax=0.56]
			\node[anchor=north east] at (axis description cs:1,1) {$\varepsilon=45\%, \mathsf{ABEL}=\frac{900}{299}$};

			\addplot+[blue,mark=,dashed] table[x=N,y=B4,col sep=comma]{data/tpM8p2l3.txt};
			\addplot+[blue,mark=,densely dotted] table[x=N,y=A4,col sep=comma]{data/tpM8p2l3.txt};
			\addplot+[blue,mark=] table[x=N,y=I4,col sep=comma]{data/tpM8p2l3.txt};

			\addplot+[red,mark=,dashed] table[x=N,y=B8,col sep=comma]{data/tpM8p2l3.txt};
			\addplot+[red,mark=,densely dotted] table[x=N,y=A8,col sep=comma]{data/tpM8p2l3.txt};
			\addplot+[red,mark=] table[x=N,y=I8,col sep=comma]{data/tpM8p2l3.txt};

			\nextgroupplot[ymin=0.19,ymax=0.46]
			\node[anchor=north east] at (axis description cs:1,1) {$\varepsilon=55\%, \mathsf{ABEL}=\frac{900}{299}$};

			\addplot+[blue,mark=,dashed] table[x=N,y=B4,col sep=comma]{data/tpM8p3l3.txt};
			\addplot+[blue,mark=,densely dotted] table[x=N,y=A4,col sep=comma]{data/tpM8p3l3.txt};
			\addplot+[blue,mark=] table[x=N,y=I4,col sep=comma]{data/tpM8p3l3.txt};

			\addplot+[red,mark=,dashed] table[x=N,y=B8,col sep=comma]{data/tpM8p3l3.txt};
			\addplot+[red,mark=,densely dotted] table[x=N,y=A8,col sep=comma]{data/tpM8p3l3.txt};
			\addplot+[red,mark=] table[x=N,y=I8,col sep=comma]{data/tpM8p3l3.txt};

		\end{groupplot}
	\end{tikzpicture}
	\caption{The normalized throughput of BNC with $M = 8$.}
	\label{fig:tpM8}
\end{figure}

The \emph{normalized throughput} of BNC at a node is the expected rank of the batches arriving at the node divided by the batch size, which is the measurement used in literature such as \cite{yang14bats,bats_book,scheduling,adaptive,ge_adaptive}.
We perform simulations to obtain the empirical throughput.
In each configuration, the source node sends $100000$ blocks of batches, where each block has $L$ batches and each batch has $M$ packets.

The plots in Figs.~\ref{fig:tpM4} and \ref{fig:tpM8} for batch sizes $M = 4$ and $8$ respectively.
In each plot, the blue and red curves correspond to block sizes $L = 4$ and $8$ respectively.
There are three curves for each $L$ in each plot.
The dashed curve, representing baseline recoding and block interleaving (BR-BI), has the lowest throughput, which is the performance of the existing interleaved minimal protocol.
The densely dotted curve, representing adaptive recoding and stream interleaving (AR-SI), has the highest throughput, which is the upper bound for the new protocol without backward compatibility.
The solid curve, which represents adaptive recoding and intrablock interleaving (AR-IBI), %
lies between the other two curves.
We only show one solid curve for each $L$ as all the dispersion efficiencies we used, as demonstrated in Table~\ref{tab:de}, give nearly the same throughput.
That is, all the curves for intrablock interleavers overlap with each other if we plot all of them.
This means that in practice, we can simply use $\effu_{-\pe_1}$ so that the computational cost is minimized.

\begin{table}
	\centering
	\scriptsize
	\caption{Normalized Throughput and Variances at the $4$-th Hop}
	\label{tab:throughput}
	\begin{tabular}{c|c|cccc|cccc}
		\multicolumn{2}{c}{} & $\effpe_{-\pe_1}$ & $\effpe_{-\pe_2}$ & $\effpe_{\ln}$ & $\effpe_{\tan^{-1}}$ & $\effu_{-\pe_1}$ & $\effu_{-\pe_2}$ & $\effu_{\ln}$ & $\effu_{\tan^{-1}}$\\
		\specialrule{.7pt}{1pt}{1pt}
		\multicolumn{10}{c}{$\varepsilon = 35\%$, $\mathsf{ABEL} = 2$}\\
		$M = 4$ & mean				& 0.44224 & 0.44308 & 0.44303 & 0.44227 & 0.44281 & 0.44247 & 0.44260 & 0.44218\\
		\cline{2-10}
		$L = 4$ & var ($\evar$)		& 13.421 & 12.917 & 13.725 & 13.819 & 13.089 & 12.259 & 13.210 & 12.217 \\
		\midrule
		$M = 4$ & mean				& 0.46362 & 0.46337 & 0.46411 & 0.46397 & 0.46382 & 0.46349 & 0.46371 & 0.46358\\
		\cline{2-10}
		$L = 8$ & var ($\evar$)		& 5.9258 & 6.2157 & 6.1090 & 6.1776 & 6.3829 & 5.7323 & 6.0450 & 5.9231 \\
		\midrule
		$M = 8$ & mean				& 0.49706 & 0.49695 & 0.49660 & 0.49711 & 0.49739 & 0.49689 & 0.49709 & 0.49716\\
		\cline{2-10}
		$L = 4$ & var ($\evar$)		& 7.1279 & 6.1029 & 7.4044 & 6.9787 & 6.7287 & 6.7590 & 6.2589 & 6.6173 \\
		\midrule
		$M = 8$ & mean				& 0.51097 & 0.51096 & 0.51099 & 0.51107 & 0.51094 & 0.51072 & 0.51096 & 0.51085\\
		\cline{2-10}
		$L = 8$ & var ($\evar$)		& 3.3326 & 3.3394 & 3.4401 & 3.5724 & 3.1228 & 3.2192 & 3.3413 & 3.2807 \\
		\specialrule{.7pt}{1pt}{1pt}

		\multicolumn{10}{c}{$\varepsilon = 45\%$, $\mathsf{ABEL} = 2.5$}\\
		$M = 4$ & mean				& 0.33351 & 0.33343 & 0.33366 & 0.33362 & 0.33349 & 0.33349 & 0.33340 & 0.33397\\
		\cline{2-10}
		$L = 4$ & var ($\evar$)		& 12.638 & 13.501 & 13.962 & 14.243 & 12.868 & 12.748 & 12.777 & 12.434\\
		\midrule
		$M = 4$ & mean				& 0.36100 & 0.36121 & 0.36123 & 0.36127 & 0.36107 & 0.36173 & 0.36128 & 0.36128\\
		\cline{2-10}
		$L = 8$ & var ($\evar$)		& 6.5622 & 6.6549 & 6.4310 & 6.5832 & 6.3443 & 6.5855 & 6.7941 & 5.7188\\
		\midrule
		$M = 8$ & mean				& 0.38693 & 0.38761 & 0.38736 & 0.38792 & 0.38789 & 0.38792 & 0.38758 & 0.38759\\
		\cline{2-10}
		$L = 4$ & var ($\evar$)		& 7.3754 & 7.7533 & 7.8033 & 7.3292 & 7.2504 & 7.3194 & 7.6881 & 7.6812\\
		\midrule
		$M = 8$ & mean				& 0.40647 & 0.40649 & 0.40676 & 0.40674 & 0.40637 & 0.40644 & 0.40634 & 0.40638\\
		\cline{2-10}
		$L = 8$ & var ($\evar$)		& 3.4554 & 3.6074 & 3.6178 & 3.6001 & 3.5397 & 3.5459 & 3.6789 & 3.6582\\
		\specialrule{.7pt}{1pt}{1pt}

		\multicolumn{10}{c}{$\varepsilon = 55\%$, $\mathsf{ABEL} = 900/299$}\\
		$M = 4$ & mean				& 0.24291 & 0.24277 & 0.24307 & 0.24331 & 0.24373 & 0.24363 & 0.24357 & 0.24310\\
		\cline{2-10}
		$L = 4$ & var ($\evar$)		& 10.153 & 10.565 & 9.8534 & 10.917 & 10.591 & 9.9482 & 10.863 & 11.403\\
		\midrule
		$M = 4$ & mean				& 0.27155 & 0.27163 & 0.27119 & 0.27173 & 0.27161 & 0.27216 & 0.27159 & 0.27155\\
		\cline{2-10}
		$L = 8$ & var ($\evar$)		& 5.8994 & 5.9115 & 5.7981 & 5.4763 & 5.3553 & 5.9013 & 5.6992 & 5.1604\\
		\midrule
		$M = 8$ & mean				& 0.29175 & 0.29202 & 0.29163 & 0.29164 & 0.29207 & 0.29202 & 0.29146 & 0.29231\\
		\cline{2-10}
		$L = 4$ & var ($\evar$)		& 6.7011 & 6.3949 & 6.5715 & 6.2801 & 6.4302 & 6.3017 & 6.3829 & 6.5296\\
		\midrule
		$M = 8$ & mean				& 0.31107 & 0.31114 & 0.31156 & 0.31129 & 0.31125 & 0.31138 & 0.31159 & 0.31123\\
		\cline{2-10}
		$L = 8$ & var ($\evar$)		& 3.3019 & 3.3151 & 3.1860 & 3.6101 & 3.4475 & 3.3164 & 3.4573 & 3.1130\\
		\bottomrule
	\end{tabular}
\end{table}

We take the $4$-th hop as an example to show how close the normalized throughput is in Table~\ref{tab:throughput}.
To further illustrate the stability of the empirical throughput, we group the empirical throughput of every $100$ blocks as a sample.
We calculated the variance of the $1000$ samples and show it in the table, which are in order $\evar$.
This small variance suggests that the throughput converges quickly and the variation of the throughput is small.

In some cases, the throughput of BR-BI is the same for different $L$.
Similar phenomenon also happens for AR-IBI.
This is related to the length of the bursts and the occurrence of the bursts.
For example, if the bursts are short but it is likely to fall in another burst after $L$ packets, then the burst is not being alleviated by packet separation in the view of most batches.
On the other hand, if the bursts are a bit long but the occurrence is small, then most batches do not have consecutive packet lost.
That is, only a small number of batches are sacrificed to have low rank but the overall rank distribution is still in a good shape, so the outer code of BNC can still achieve a relatively high rate.
In other words, our simulations suggest that in order to gain a greater benefit of using interleavers, the statistics of average burst error length is not enough for determining the interleaver depth or the block size.
Yet, the choice of $L$ is out of the scope of this paper, but we remark that the choice of $L$ also affects the latency.
For example, if we generate and transmit recoded packets after all the packets in a block which are not lost are received, then the latency is $ML$ packets.

In general, we can see that the throughput is higher when $L$ is larger and/or $M$ is larger.
The throughput gain of AR-IBI using a larger $L$ is significant in many cases, and sometimes the throughput of AR-IBI is close to the one of AR-SI when $L = 8$.
Nevertheless, we can observe a significant throughput gain of AR-IBI comparing with BR-BI.

\section{Concluding Remarks}
\label{sec:conclude}

The essential problem studied in this paper is the block-by-block interleaving of batches.
When all the batches have the same number of packets, block interleaving is commonly applied and supposed to be the optimal interleaving strategy.
However, when the batches may have different number of packets, the problem, as far as we know, has not been studied in literature.
Borrowing the idea from physics, we use potential energy in classical mechanics to measure the performance of an interleaver, and discussed the economy interpretation of the measurement.
We also proposed an algorithm to optimize the interleaver with this performance measure.
Our solution here may of general interest for interleaving research. 

For batched network coding, the intrablock interleaving strategy is of interest in the design of packet scheduling when the global statistics of the batches are not known in advance, as a block acts as a window for a short observation of the recent channel condition.
This strategy also bounds the latency and buffer size.
Our solution here enables the joint optimization of adaptive recoding and intrablock interleaving.
This combination of interleaving and adaptive recoding can not only gain the advantage of both, but also remove the implementation issues of stream interleaving.
As one of the advantages of our solution, for existing systems which use baseline recoding and block interleaving, upgrading to adaptive recoding and intrablock interleaving are much more straightforward than using adaptive recoding and stream interleaving.
It is even possible to allow some nodes use adaptive recoding and intrablock interleaving, while other nodes use baseline recoding and block interleaving.
This compatibility can enhance the performance of existing systems without overhauling the protocol architecture, which is a feasible solution in case some devices are deployed at a location which cannot be easily reached.

As a future research, the trade-off between the throughput gained and the latency induced by interleaving is of interest.
For example, we may study the variations of stream interleaving with bounded latency between the first and the last packets.
We are curious the performance compared with the intrablock interleaver. 
On the other hand, our numerical evaluations suggested that the statistics of average burst error length is not enough for determining an interleaver depth or a block size which can guarantee a good alleviating ability of rank loss within a batch due to burst packet loss.
A deeper understanding of the relation between burst loss models and batched network codes is another research direction.

%

%
%
%
%

%
%

%

%

%

%

%

%

%

%

\newpage

\appendices

\section{Evaluations of the Gilbert-Elliott Model}
\label{sec:ge}

As a reasonable GE model, we assume that $0 < p, q < 1$.

\subsection{Expected Rank Functions}
\label{sec:ge:exp}

The part related to the channel model in the expected rank functions is the terms $\Pr(X_{t_k} = i)$ for all $i = 0, 1, \ldots, t_k$.
When we know how to formulate $\Pr(X_{t_k} = i)$, we can calculate the expected rank functions.
By exploiting the structure of the GE model, we can calculate $\Pr(X_{t_k} = i)$ from a given $\mathcal{S}_k$.

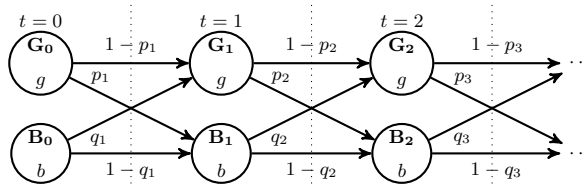
\begin{figure}
	\centering
	\begin{tikzpicture}[scale=0.8,font=\footnotesize,state/.style={circle,draw=black,thick,inner sep=1pt,minimum size=20pt,align=center},->,>=stealth',thick,every node/.style={transform shape}]
		\foreach \i in {0,...,2}{
			\begin{scope}[shift={(3*\i, 0)}]
				\node[state] (G\i) at (0,1.5) {$\mathbf{G_{\i}}$\\$g$};
				\node[state] (B\i) at (0,0) {$\mathbf{B_{\i}}$\\$b$};
				\node at (0,2.2) {$t = \i$};
				\draw[-,thin,dotted] (1.5,-.5) -- ++(0,3);
			\end{scope}
		}
		\node[circle,thick,inner sep=1pt,minimum size=20pt] (G3) at (9,1.5) {$\cdots$};
		\node[circle,thick,inner sep=1pt,minimum size=20pt] (B3) at (9,0) {$\cdots$};
		\foreach \i in {0,...,2}{
			\pgfmathtruncatemacro{\j}{\i+1};
			\path (G\i) edge node[above,pos=.25] {$p_{\j}$} (B\j);
			\path (B\i) edge node[below,pos=.25] {$q_{\j}$} (G\j);
			\path (G\i) edge node[above] {$1-p_{\j}$} (G\j);
			\path (B\i) edge node[below] {$1-q_{\j}$} (B\j);
		}
	\end{tikzpicture}
	\caption{The Gilbert-Elliott model when the transition matrices are different for different transmissions.}
	\label{fig:long_chain}
\end{figure}

First, the transition matrix of the Markov chain of the GE model is
\begin{equation*}
	\mathbf{P} = \begin{pmatrix}
		1-p & p\\
		q & 1-q
	\end{pmatrix},
\end{equation*}
with the stationary distribution $(\pi_\mathbf{G}, \pi_\mathbf{B}) = (\frac{q}{p+q}, \frac{p}{p+q})$.
Next, the IPG between the $j$-th and $(j+1)$-th packets is $x_{k,j+1}-x_{k,j}$, which is equivalent to run the Markov chain $x_{k,j+1}-x_{k,j}$ times in a row when we change the state of the chain for sending the $(j+1)$-th packet.
Let
\begin{equation*}
	\mathbf{P}_{k,j+1} :=
	\begin{pmatrix}
		1-p_{k,j+1} & p_{k,j+1}\\
		q_{k,j+1} & 1-q_{k,j+1}
	\end{pmatrix}
\end{equation*}
be the transition matrix for transmitting the $(j+1)$-th packet of the batch.
For $j \ge 1$, we have $\mathbf{P}_{k,j+1} = \mathbf{P}^{x_{k,j+1}-x_{k,j}}$.
On the other hand, we have $\mathbf{P}_{k,1} = \mathbf{P}^{x_{k,1}}$ as we run the GE model from the $1$-st packet in the TS.
We can interpret these transition matrices by duplicating the GE model to form a longer Markov chain as illustrated in Fig.~\ref{fig:long_chain}.
The initial state is either $\mathbf{G_0}$ or $\mathbf{B_0}$ with probabilities %
$(\pi_\mathbf{G}, \pi_\mathbf{B})$.

Let $\mathfrak{S}_{k,t}$ be the $\{\mathbf{G}, \mathbf{B}\}$-valued random variables of the state of the Markov chain after transmitting the $t$-th packet of the batch $b_k$.
For simplicity, define $\Pr(\mathfrak{s}, i, t_k) := \Pr(\mathfrak{S}_{k,t} = \mathfrak{s}, X_{t_k} = i)$.
Then, we have
\begin{equation*}
	\Pr(X_{t_k} = i) = \Pr(\mathbf{G}, i, t_k) + \Pr(\mathbf{B}, i, t_k).
\end{equation*}

By exploiting the structure of the GE model, we have the recursive relation
\begin{multline*} %
	\begin{pmatrix}
		\Pr(\mathbf{G}, i, j+1)\\
		\Pr(\mathbf{B}, i, j+1)
	\end{pmatrix}
	=
	\begin{pmatrix}
		g & 0\\
		0 & b
	\end{pmatrix}
	(\mathbf{P}_{k,j+1})^\intercal
	\begin{pmatrix}
		\Pr(\mathbf{G}, i, j)\\
		\Pr(\mathbf{B}, i, j)
	\end{pmatrix}\\
	+
	\begin{pmatrix}
		1-g & 0\\
		0 & 1-b
	\end{pmatrix}
	(\mathbf{P}_{k,j+1})^\intercal
	\begin{pmatrix}
		\Pr(\mathbf{G}, i-1, j)\\
		\Pr(\mathbf{B}, i-1, j)
	\end{pmatrix},
\end{multline*}
where the boundary conditions are
\begin{enumerate}[a)]
	\item 
		$\Pr(\cdot, i, j) = 0$ for $i < 0$ or $i > j$; and
	\item 
		$\Pr(\mathbf{G}, 0, 0) = \pi_\mathbf{G}$ and $\Pr(\mathbf{B}, 0, 0) = \pi_\mathbf{B}$.
\end{enumerate}

The value of %
$\Pr(X_{t_k} = i)$, and thus $E(r_k, \mathcal{S}_k)$,
can be evaluated by dynamic programming.
Note that the power $x_{k,1}$ in $\mathbf{P}^{x_{k,1}}$ is not important during the evaluation because we have $(\pi_\mathbf{G}, \pi_\mathbf{B})\mathbf{P}^{x_{k,1}} = (\pi_\mathbf{G}, \pi_\mathbf{B})$.
In other words, we can shift the indices in $\mathcal{S}_k$ to have $x_{k,1} = 1$, i.e., we have $E(r_k,\mathcal{S}_k) = E(r_k,\mathcal{S}_k-x_{k,1}+1)$.

\subsection{Pseudo Interleaver Depths}
\label{sec:ge:pid}

Now, we discuss how to obtain the new channel models and the pseudo interleaver depths.
For a Markov chain, the short-term transition matrix can be regarded as the root of the transition matrix for a longer observation \cite{credit,transition},
although the non-negative power of an arbitrary stochastic matrix may not be stochastic \cite{pth}.
The transition matrix we can observe after the node transmits the last packet of $b_k$ is
\begin{equation*}
	\prod_{i = 1}^{t_k} \mathbf{P}_{k,i} = \mathbf{P}^N
\end{equation*}
for some non-negative integer $N$.
When $t_k = 1$, this single recoded packet can be located anywhere in the block with no concept of spreading. %
We can see from the above equation that $N = 1$ for this case.
Now consider $t_k > 1$.
By taking the average, we consider that each transmission of the batch takes the same number of steps on the GE model.
So, the transition matrix for each transmission is the ``geometric mean'', i.e., the $t_k$-th root, of $\mathbf{P}^N$. %
The pseudo interleaver depth $L_k$ is thus $N/t_k$.

Note that $\mathbf{P}$ is a diagonalizable matrix.
Let $\mathbf{P} = \mathbf{V}\mathbf{D}\mathbf{V}^{-1}$ for some invertible matrix $\mathbf{V}$ and diagonal matrix $\mathbf{D}$.
A possible candidate is
\begin{equation*}
	\mathbf{V} = \begin{pmatrix}
		1 & \frac{-p}{q}\\
		1 & 1
	\end{pmatrix} \quad \text{and} \quad
	\mathbf{D} = \begin{pmatrix}
		1 & 0 \\
		0 & -p-q+1
	\end{pmatrix}.
\end{equation*}
The definition of non-negative power, say $N/t_k$, of the diagonalizable matrix $\mathbf{P}$ is $\mathbf{V}\mathbf{D}^{N/t_k}\mathbf{V}^{-1}$.
We can verify that
\begin{equation*}
	\mathbf{P}^{N/t_k} = \begin{pmatrix}
		\frac{q + p(-p-q+1)^{N/t_k}}{p+q} & \frac{p - p(-p-q+1)^{N/t_k}}{p+q}\\
		\frac{q - q(-p-q+1)^{N/t_k}}{p+q} & \frac{p + q(-p-q+1)^{N/t_k}}{p+q}
	\end{pmatrix}.
\end{equation*}
is a stochastic matrix. %
In other words, we consider a GE model with a transition matrix $\mathbf{P}^{N/t_k}$ for the recoded packets of the $k$-th batch.
Yet, we need to find a reasonable $N$.
For the indices stated in $s^{-1}(k)$, we have $\mathbf{P}_{k,j+1} = \mathbf{P}^{x_{k,j+1}-x_{k,j}}$. %
By telescoping, we have
\begin{equation*}
	\prod_{i = 2}^{t_k} \mathbf{P}_{k,i} = \mathbf{P}^{x_{k,t_k}-x_{k,1}}.
\end{equation*}
Recall that $\mathbf{P}_{k,1}$ can be an arbitrary power of $\mathbf{P}$.
As the geometric mean of the transition matrices simulates the channel condition for this batch where the IPGs between the packets of this batch are being averaged, we should follow this condition for the $1$-st packet so that everything is captured by the new GE model.
In other words, we have
\begin{equation*}
	N = \frac{t_k(x_{k,t_k}-x_{k,1})}{t_k-1}, t_k > 1.
\end{equation*}

When $t_k = 0$, we have no information on how to spread the packets of this batch.
We can simply set $L_k = 1$ so it would follow the original channel condition if we are going to send some packets of this batch.
Combining all the cases, we conclude that the transition matrix of the new GE model for the $k$-th batch is $\mathbf{P}^{L_k}$ with a pseudo interleaver depth 
\begin{equation*}
	L_k := \begin{cases}
		(x_{k,t_k}-x_{k,1})/(t_k-1) & \text{if } t_k > 1,\\
		1 & \text{otherwise}.
	\end{cases}
\end{equation*}

We remark that the above method to calculate the non-negative power of the transition matrix can be used to synchronize the timescale.

\subsection{Burst Error Length}
\label{sec:ge:burst}

For a simplified GE model where $g = 0$ and $b = 1$, the length of burst error follows a geometric distribution so that we know the ABEL is $1/q$ as suggested in \cite{hasslinger08}.
For a general GE model, we need a bit more effort to obtain the distribution of burst error length.

The probability of the first error in the burst occurs in each state $\mathbf{S} \in \{\mathbf{G}, \mathbf{B}\}$ is denoted by $\wp_\mathbf{S}$.
The probability of reaching state $\mathbf{G}$ and having an error while there is no error in the previous state is 
\begin{equation*}
	\widetilde{\wp_\mathbf{G}} := (1-p)g(1-g)\pi_\mathbf{G} + qg(1-b)\pi_\mathbf{B} = \frac{qg((1-g)-p(b-g))}{p+q}.
\end{equation*}
The one for state $\mathbf{B}$ is
\begin{equation*}
	\widetilde{\wp_\mathbf{B}} := pb(1-g)\pi_\mathbf{G} + (1-q)b(1-b)\pi_\mathbf{B} = \frac{pb((1-b)+q(b-g))}{p+q}.
\end{equation*}
As they are the only possibilities of starting a new burst, we have
\begin{equation*}
	\wp_\mathbf{G} = \frac{\widetilde{\wp_\mathbf{G}}}{\widetilde{\wp_\mathbf{G}}+\widetilde{\wp_\mathbf{B}}},
	\quad
	\wp_\mathbf{B} = \frac{\widetilde{\wp_\mathbf{B}}}{\widetilde{\wp_\mathbf{G}}+\widetilde{\wp_\mathbf{B}}}.
\end{equation*}

Let $P_{\mathbf{S},i}$ be the probability of having the $i$-th consecutive errors in state $\mathbf{S} \in \{\mathbf{G}, \mathbf{B}\}$.
We can model a recursive relation for having one more error from $i$ consecutive errors by
\begin{equation*}
	P_{\mathbf{G},i+1} = (1-p)g P_{\mathbf{G},i} + qg P_{\mathbf{B},i},
	\quad 
	P_{\mathbf{B},i+1} = pb P_{\mathbf{G},i} + (1-q)b P_{\mathbf{B},i}.
\end{equation*}
We can see that the initial conditions are $P_{\mathbf{G},1} = \wp_\mathbf{G}$ and $P_{\mathbf{B},1} = \wp_\mathbf{B}$.

We now express the average burst error length (ABEL), denoted by $\mathsf{ABEL}$. %
Let $e_\mathbf{S}$ be the ABEL where the first error in the burst occurs in state $\mathbf{S} \in \{\mathbf{G}, \mathbf{B}\}$.
The probability to stay in \textbf{G} and lose a packet is $(1-p)g$, the one to transit from \textbf{G} to \textbf{B} and lose a packet is $pb$, etc.
Thus,
\begin{equation*}
	\begin{pmatrix}
		e_\mathbf{G}\\
		e_\mathbf{B}
	\end{pmatrix} = \begin{pmatrix}
		(1-p)g & pb\\
		(1-q)b & qg
	\end{pmatrix}\begin{pmatrix}
		e_\mathbf{G}\\
		e_\mathbf{B}
	\end{pmatrix} + \begin{pmatrix}
		1\\
		1
	\end{pmatrix}.
\end{equation*}
We can view this formulation as adding an absorbing state for the termination of a burst.
As the expected stopping time to reach this new state from itself is $0$, we can remove the row and column for this state from the transition matrix and obtain the above $2 \times 2$ matrix.
Therefore, by the law of total expectation, %
we have
\begin{equation*}
	\mathsf{ABEL} = \wp_\mathbf{G} e_\mathbf{G} + \wp_\mathbf{B} e_\mathbf{B}.
\end{equation*}

Lastly, we want to find the transition probabilities in terms of average loss rate and ABEL.
The average loss rate, denoted by $\varepsilon$, equals $g \pi_\mathbf{G} + b \pi_\mathbf{B} = \frac{gq+bp}{p+q}$.
That is, we have $q (\varepsilon-g) = p (b-\varepsilon)$.
We can see that if $\varepsilon$ equals either $g$ or $b$, then we have $g = b$, and arbitrary $0 < p, q < 1$ give the desired average loss rate.
Here, we assume $\varepsilon \neq g, b$.
Then, we have
\begin{equation} \label{eq:q}
	q = \frac{p(b-\varepsilon)}{\varepsilon-g}.
\end{equation}
We can then substitute \eqref{eq:q} into the formula of ABEL and change the subject to $p$. %
The arithmetic steps are tedious and we leave the derivation in Appendix~\ref{sec:ge2}.
Finally, we obtain a quadratic equation $Ap^2 + Bp + C = 0$ where
\begin{IEEEeqnarray*}{rCl}
	A & = & \mathsf{ABEL}(b-\varepsilon)(b-g)^3((1-b)(1-g)-(1-\varepsilon))\\
	B & = & \mathsf{ABEL}(b-g)[((1-b)(1-g)-(1-\varepsilon))(g(1-g)(b-\varepsilon)+b(1-b)(\varepsilon-g))\\
	&& \quad -(b-\varepsilon)(\varepsilon-g)(1-b)(1-g)(b-g)] - \varepsilon(b-g)^2((1-b)(1-g)-(1-\varepsilon))\\
	C & = & (\varepsilon-g)(1-b)(1-g)[\varepsilon(b-g)-\mathsf{ABEL}(g(1-g)(b-\varepsilon)+b(1-b)(\varepsilon-g))].
\end{IEEEeqnarray*}
By selecting a root $0 < p < 1$, we can obtain $q$ by substituting $p$ into \eqref{eq:q}.
We remark that under a fixed $b$ and $g$, a GE model may not be able to represent an arbitrary $\varepsilon$ and $\mathsf{ABEL}$, i.e., the above quadratic equation may not have a root $p \in (0, 1)$.

\section{Total Potential Energy and Particle Separation}
\label{sec:potential}

Suppose we have $n+1$ identical particles.
We fix two particles and put the other $n-1$ particles between them colinearly, and sequentially call them the $1$-st particle, the $2$-nd particle, and so on.
Let $r_i$ be the distance between the $i$-th and $(i+1)$-th particles for $i = 1, 2, \ldots, n$.
For simplicity, let $\sum_{i = 1}^n r_i = c$ for some positive $c$.
As discussed in Sec.~\ref{sec:pe_model}, we consider repulsive interactions only, and then the potential energy of a particle induced by another particle, $\pe(r)$, is a strictly decreasing convex function of the distance between the two particles.

When we only consider the interactions between neighboring particles as in the Ising model, the particles are uniformly separated, i.e., $r_i$ are all equal for all $i$.
We justify this phenomenon as follows.
The potential energy minimization problem is
\begin{equation} \label{eq:ising_pe}
	\min_{r_i, \forall i} \sum_{i = 1}^n \pe(r_i) \quad \text{s.t.} \quad \sum_{i = 1}^n r_i = c.
\end{equation}
By Jensen's inequality, we have
\begin{equation*}
	\sum_{i = 1}^n \pe(r_i) \ge n \pe\left(\sum_{i = 1}^n r_i/n\right) = n \pe(c/n).
\end{equation*}
The equality holds when $r_i = c/n$ for all $i$.
In other words, $r_i = c/n$ for all $i$ can solve \eqref{eq:ising_pe}, which verifies the phenomenon of uniform particle separation.

However, when we consider the interactions between all pairs of particles, we would observe a perturbation from a uniform particle separation.
Take $n+1 = 4$ as an example.
The $1$-st particle pushes the $2$-nd particle away from it, while the $3$-rd and the $4$-th particles push the $2$-nd particle to the opposite side.
In other words, in an equilibrium state, the $2$-nd particle should be closer to the $1$-st particle, i.e., $r_1 < c/3$.
By symmetry, we have $r_3 = r_1 < c/3$ and thus $r_2 > c/3$.
To illustrate how large the perturbation is, we take $\pe(r) = 1/r$, let $r_1 = r_3 = c/3 - x$ and let $r_2 = c/3 + 2x$.
The potential energy minimization problem is
\begin{IEEEeqnarray*}{Cl}
	& \min_{x \in (0, c/3)} \frac{1}{r_1} + \frac{1}{r_2} + \frac{1}{r_3} + \frac{1}{r_1+r_2} + \frac{1}{r_2+r_3} + \frac{1}{r_1+r_2+r_3}\\
	= & \min_{x \in (0, c/3)} \frac{2}{c/3-x} + \frac{1}{c/3+2x} + \frac{2}{2c/3+x} + \frac{1}{c}
\end{IEEEeqnarray*}
By derivative test, we know that the minimum is reached for an $x \in (0, c/3)$ satisfying the quartic equation $81x^4 - 594cx^3 - 351c^2 x^2 - 66c^3 x + c^4 = 0$.
The roots are $\frac{c}{3} \left( \frac{11}{2} + \frac{9}{\sqrt{2}} \pm \frac{3}{2}\sqrt{31+22\sqrt{2}} \right)$ and $\frac{c}{6} \left( 11 - 9\sqrt{2} \pm 3 \sqrt{-1} \sqrt{22\sqrt{2}-31} \right)$.
Only $x = \frac{c}{3} \left( \frac{11}{2} + \frac{9}{\sqrt{2}} - \frac{3}{2}\sqrt{31+22\sqrt{2}} \right)$ is a real root in $(0, c/3)$. %
That is, we have $r_1 = r_3 \approx 0.3193c$ and $r_2 \approx 0.3615c$. %
In the view of TS, we force all $r_i$ to be integers and we also have $c \ge n = 3$.
If $c$ is small, i.e., the uniformness of the separation matters, the perturbation is likely to be eliminated after we round $r_i$ to integers.
If $c$ is large, i.e., the separation is already very large, the perturbation barely affects the dilution of burst errors.
From our evaluations in Sec.~\ref{sec:sim}, we do not observe a drop in throughput due to the perturbation.
Despite of this, we recommend to consider the interactions between neighboring particles only to reduce computation.

\section{Benchmark Generation}

\subsection{Simulated Annealing}
\label{sec:sa}

Simulated annealing (SA) \cite{sa}, an analogue of annealing in metallurgy, is a well-known probabilistic technique for approximating optimization problems with large search space.
We briefly describe the algorithm here.
Each item in the search space is regarded as a state.
The states which can be reached from a state in one transition are called the neighbours of the state.
A neighbour function would output a neighbour of the given state randomly.
Each state corresponds to a certain amount of energy.
The energy function outputs the energy of a given state.
The algorithm starts from a given initial state and a given initial temperature.
In each iteration, the algorithm queries the neighbour function for a neighbour of the current state.
If the selected neighbour has a smaller energy, then the algorithm would transit to this neighbour.
Otherwise, there is a chance to transit to this neighbour according to the acceptance probability.
The acceptance probability is a function in terms of the current temperature, the energy of the current state and the energy of the selected neighbour.
The temperature is dropped after each iteration.
Additionally, we add an extra step to the algorithm which records the lowest energy among the states travelled by the algorithm.

In order to apply SA to our problem, we have the following parameters and settings.
Each TS $f \in \mathcal{F}$ is regarded as a state.
Although we can run SA from a state which is not given by the above algorithms, the outcome is better if we do so because we are making use of the knowledge of the optimization problem.
As the SA algorithm tries to minimize the energy, our energy function is the negation of the dispersion efficiency, i.e., $-\eff(f)$.
The acceptance probability we adopted is the Boltzmann factor $e^{\frac{-\Delta E}{kT}}$ with $k = 1$ where $T$ is the current temperature and $\Delta E$ is the change of energy functions from the current state to a new state.
The neighbour function swaps two components in the TS $f$ where the swapped components have distinct values, i.e., the function generates a TS $f'$ from $f$ where the Hamming distance between $f'$ and $f$ is exactly $2$.
This way, we can prevent the search being trapped around the local optimum near the initial state.
In our evaluations, we start with an initial temperature $5000$, cool down $5\%$ of the temperature after each iteration, and terminate the process when the temperature is no more than $0.0001$.
\subsection{Constraint Programming}
\label{sec:cp}

To verify the quality of the TS generated by our algorithms, we also model the problem in the constraint programming (CP) \cite{cp} framework and try to find the exact optimum.
We apply various techniques including symmetry breaking \cite{symmetry} and tabulation with element constraint \cite{elementconst} to improve the solving efficiency.
Despite the application of the above techniques, it still takes a long time to find an optimal TS, so we only apply CP on some samples for comparison only.

The idea of symmetry breaking is as follows.
Note that the system energy is the sum of the potential energies of the individual batches.
If $t_i = t_j$, namely, the number of packets for the $i$-batch and the $j$-batch are the same, then exchange the positions of packets of the $i$-batch and those of the $j$-batch will not change the value of the potential energy.
Therefore, we impose the constraint that the position of the first packet of the $i$-batch must be less than that of $j$-batch, i.e., $x_{1,i} < x_{1, j}$, if $t_i = t_j$ and $i < j$.

Regarding tabulation, note that the distance $r$ in the potential energy is the output of the metric $d \colon \mathcal{M}^2 \to \mathbb{R}$.
On the other hand, CP is a paradigm for solving discrete optimization problems.
The presence of floating point arithmetic in the objective may slow down the solving efficiency of a CP solver.
Therefore, we tabulate the objective function and use the element constraints to retrieve the objective value.
The constructed table $obj$ is an $\mathcal{M} \times \mathcal{M}$ $2$-D array, where each element $obj[i, j]$ is the value of the potential energy $\pe_n(i,j)$.
Thus, the potential energy objective in (\ref{eq:PE}) becomes 
\begin{equation*}
\min_{x_{k,i}, \forall k, i} \quad \sum_{k = 1}^L \sum_{i = 1}^{t_k-1} \sum_{j = i+1}^{t_k} obj[x_{k,j}, x_{k,i}]\\
\end{equation*}

The model has been written in a high-level modelling language MiniZinc~\cite{minizinc}, and the compiler can convert the model into FlatZinc, a solver input language that is understood by various constraint satisfaction solvers such as Gecode~\cite{gecode} and COIN-OR BC solver~\cite{coin-bc}. We choose to use COIN-OR BC, a mixed integer programming solver, as the low-level solver, and the compiler of MiniZinc can apply linearization~\cite{linearization,linearization2} and automatically covert a high-level non-linear model into a linear one.

\section{Deriving the Transition Probability}
\label{sec:ge2}

We need to write the formula of $\mathsf{ABEL}$ explicitly before we can change the subject of the expression to $p$.
First, we have
\begin{equation*}
	\wp_\mathbf{G} = \frac{qg((1-g)-p(b-g))}{qg(1-g)+pb(1-b)+pq(b-g)^2},
	\quad
	\wp_\mathbf{B} = \frac{pb((1-b)+q(b-g))}{qg(1-g)+pb(1-b)+pq(b-g)^2}.
\end{equation*}
Next, we solve the system of linear equations of $e_\mathbf{G}$ and $e_\mathbf{B}$ and obtain 
\begin{equation*}
	\begin{pmatrix}
		e_\mathbf{G}\\
		e_\mathbf{B}
	\end{pmatrix} = \frac{1}{b(1-q)+g(1-p)-bg(1-p-q)-1}
	\begin{pmatrix}
		(1-p-q)b-1\\
		(1-p-q)g-1
	\end{pmatrix}.
\end{equation*}
Then, we can write
\begin{multline} \label{eq:fullL}
	\mathsf{ABEL} = \wp_\mathbf{G} e_\mathbf{G} + \wp_\mathbf{B} e_\mathbf{B}\\
	= \frac{qg((1-g)-p(b-g))((1-p-q)b-1)+pb((1-b)+q(b-g))((1-p-q)g-1)}{(qg(1-g)+pb(1-b)+pq(b-g)^2)(b(1-q)+g(1-p)-bg(1-p-q)-1)}.
\end{multline}

In the following, we substitute
	$q = \frac{p(b-\varepsilon)}{\varepsilon-g}$
into \eqref{eq:fullL}. %
Note that
\begin{equation*}
	1-p-q = 1-p-\frac{p(b-\varepsilon)}{\varepsilon-g} = \frac{(1-p)(\varepsilon-g)-p(b-\varepsilon)}{\varepsilon-g} = \frac{(\varepsilon-g) - p(b-g)}{\varepsilon-g}.
\end{equation*}

We start from the numerator of $\mathsf{ABEL}$ in \eqref{eq:fullL} times $\frac{(\varepsilon-g)^2}{p}$.
\begin{IEEEeqnarray*}{Cl}
	& \frac{(\varepsilon-g)^2}{p}[qg((1-g)-p(b-g))((1-p-q)b-1)+pb((1-b)+q(b-g))((1-p-q)g-1)]\\
	= & (g(1-g)(b-\varepsilon)-pg(b-g)(b-\varepsilon)) \overbrace{\left( \frac{(\varepsilon-g) - p(b-g)}{\varepsilon-g} b -1 \right)(\varepsilon-g))}^{= b(\varepsilon-g)-pb(b-g)-(\varepsilon-g) = -(\varepsilon-g)(1-b)-pb(b-g)}\\
	& \quad + (b(1-b)(\varepsilon-g)+pb(b-g)(b-\varepsilon)) \underbrace{\left( \frac{(\varepsilon-g) - p(b-g)}{\varepsilon-g} g - 1 \right)(\varepsilon-g)}_{= g(\varepsilon-g)-pg(b-g)-(\varepsilon-g) = -(\varepsilon-g)(1-g)-pg(b-g)}\\
	= & -g(1-g)(b-\varepsilon)(\varepsilon-g)(1-b) + p [ -g(1-g)(b-\varepsilon)b(b-g) + g(b-g)(b-\varepsilon)(\varepsilon-g)(1-b) ]\\
	& \quad + \cancel{p^2 gb (b-g)^2(b-\varepsilon)}\\
	& \qquad - b(1-b)(\varepsilon-g)^2(1-g) + p [ -b(1-b)(\varepsilon-g)g(b-g) - b(b-g)(b-\varepsilon)(\varepsilon-g)(1-g) ]\\
	& \qquad\quad - \cancel{p^2 gb (b-g)^2(b-\varepsilon)}\\
	= & -(1-g)(1-b)(\varepsilon-g) \overbrace{(g(b-\varepsilon)+b(\varepsilon-g))}^{= \varepsilon(b-g)} + p(b-g)\\
	& \quad \times [-bg(1-g)(b-\varepsilon)+g(b-\varepsilon)(\varepsilon-g)(1-b)-b(b-\varepsilon)(\varepsilon-g)(1-g)-bg(1-b)(\varepsilon-g)]\\
	= & -\varepsilon(1-g)(1-b)(\varepsilon-g)(b-g) + p(b-g)\\
	& \quad \times [ (b-\varepsilon)(\varepsilon-g)\underbrace{(g(1-b)-b(1-g))}_{= -(b-g)} - bg\underbrace{((1-g)(b-\varepsilon) + (1-b)(\varepsilon-g))}_{= b + g\varepsilon - g - b\varepsilon = (b-g)(1-\varepsilon)} ]\\
	= & -\varepsilon(1-g)(1-b)(\varepsilon-g)(b-g) - p (b-g)^2 \overbrace{((b-\varepsilon)(\varepsilon-g) + bg(1-\varepsilon))}^{=\varepsilon(b+g-bg-\varepsilon) = \varepsilon((1-\varepsilon)-(1-b)(1-g))}\\
	= & -\varepsilon(1-g)(1-b)(\varepsilon-g)(b-g) + p \varepsilon (b-g)^2 ((1-b)(1-g)-(1-\varepsilon)).
\end{IEEEeqnarray*}

Now, we consider the denominator of $\mathsf{ABEL}$ in \eqref{eq:fullL} times $\frac{(\varepsilon-g)^2}{p}$.
\begin{IEEEeqnarray*}{Cl}
	& \frac{(\varepsilon-g)^2}{p}(qg(1-g)+pb(1-b)+pq(b-g)^2)(b(1-q)+g(1-p)-bg(1-p-q)-1)\\
	= & (g(1-g)(b-\varepsilon)+b(1-b)(\varepsilon-g)+p(b-\varepsilon)(b-g)^2)\\
	& \quad \left( \frac{(\varepsilon-g) - p(b-g)}{\varepsilon-g} b + g(1-p) - \frac{(\varepsilon-g) - p(b-g)}{\varepsilon-g} bg - 1 \right) (\varepsilon-g)\\
	= & (g(1-g)(b-\varepsilon)+b(1-b)(\varepsilon-g)+p(b-\varepsilon)(b-g)^2)\\
	& \quad (b(\varepsilon-g) - bp(b-\varepsilon) + g(\varepsilon-g)(1-p) - bg(\varepsilon-g) + bgp(b-g) - (\varepsilon-g))\\
	= & (g(1-g)(b-\varepsilon)+b(1-b)(\varepsilon-g)+p(b-\varepsilon)(b-g)^2)\\
	& \quad (b(\varepsilon-g) + g(\varepsilon-g) - bg(\varepsilon-g) - (\varepsilon-g) + p\underbrace{(-b(b-\varepsilon) - g(\varepsilon-g) + bg(b-g))}_{\substack{= \varepsilon(b-g) + bg(b-g) - (b+g)(b-g)\\= (b-g)((1-b)(1-g)-(1-\varepsilon))}})\\
	= & -(g(1-g)(b-\varepsilon)+b(1-b)(\varepsilon-g))(\varepsilon-g)(1-b)(1-g)\\
	& \quad + p(b-g) [ ((1-b)(1-g)-(1-\varepsilon))(g(1-g)(b-\varepsilon)+b(1-b)(\varepsilon-g))\\
	& \qquad -(b-\varepsilon)(\varepsilon-g)(1-b)(1-g)(b-g) ]\\
	& \qquad\quad p^2 (b-\varepsilon)(b-g)^3((1-b)(1-g)-(1-\varepsilon)).
\end{IEEEeqnarray*}

The last step is to put the numerator and denominator together, and change to subject to $p$.
Then, we obtain
\begin{IEEEeqnarray*}{l}
	p^2 \mathsf{ABEL}(b-\varepsilon)(b-g)^3((1-b)(1-g)-(1-\varepsilon))\\
	\quad + p \{ \mathsf{ABEL}(b-g)[((1-b)(1-g)-(1-\varepsilon))(g(1-g)(b-\varepsilon)+b(1-b)(\varepsilon-g))\\
	\qquad -(b-\varepsilon)(\varepsilon-g)(1-b)(1-g)(b-g)] - \varepsilon(b-g)^2((1-b)(1-g)-(1-\varepsilon)) \}\\
	\qquad\quad + (\varepsilon-g)(1-b)(1-g)[\varepsilon(b-g)-\mathsf{ABEL}(g(1-g)(b-\varepsilon)+b(1-b)(\varepsilon-g))] = 0.
\end{IEEEeqnarray*}

\end{document}